\documentclass[twocolumn]{aastex62}
\usepackage{natbib}

\newcommand{\swift}{\textit{Swift}}
\newcommand{\xmm}{\textit{XMM-Newton}}
\newcommand{\inte}{\textit{INTEGRAL}}

\newcommand{\src}{{IGR\,J18027-2016}}
\newcommand{\nh}{N$_{\rm H}$}
\def \ferg {erg cm$^{-2}$ s$^{-1}$}
\def \lerg{{\rm erg~s$^{-1}$}}
\begin{document}

\title{Probing clumpy wind accretion in \src\ with XMM-Newton}
\author{Pragati Pradhan}\thanks{pragati2707@gmail.com}
\affiliation{Department of Astronomy and Astrophysics, Pennnsylvania State University, Pennnsylvania, 16802, US}
\author{Enrico Bozzo}
\affiliation{University of Geneva, Department of Astronomy, Chemin d'Ecogia 16, 1290, Versoix, Switzerland}
\author{Biswajit Paul}
\affiliation{Raman Research Institute, Astronomy and Astrophysics, C. V. Raman Avenue, Bangalore 560080. Karnataka India}
\author{Antonis Manousakis}
\affiliation{Sharjah Academy of Astronomy, Space Sciences, and Technology (SAASST) and Department of Applied Physics and Astronomy, University of Sharjah, P.O. Box 27272, Sharjah, United Arab Emirates. }
\author{Carlo Ferrigno}
\affiliation{University of Geneva, Department of Astronomy, Chemin d'Ecogia 16, 1290, Versoix, Switzerland}

\shorttitle{Clumpy wind accretion in \src\ }
\shortauthors{P. Pradhan et al.}

\begin{abstract}
Supergiant X-ray binaries usually comprise a neutron star accreting from the wind of a OB supergiant companion. They are classified as classical systems and the supergiant fast X-ray transients (SFXTs). The different behavior of these sub-classes of sources in X-rays, with SFXTs displaying much more pronounced variability, is usually (at least) partly ascribed to different physical properties of the massive star clumpy stellar wind. In case of SFXTs, a systematic investigation of the effects of clumps on flares/outbursts of these sources has been reported by \citet{bozzo2017} exploiting the capabilities of the instruments on-board XMM-Newton to perform a hardness-resolved spectral analysis on timescales as short as a few hundreds of seconds. In this paper, we use six XMM-Newton observations of IGR J18027-2016 to extend the above study to a classical supergiant X-ray binary and compare the findings with those derived in the case of SFXTs.  As these observations of IGR J18027-2016 span different orbital phases, we also study its X-ray spectral variability on longer timescales and compare our results with previous publications.  Although obtaining measurements of the clump physical properties from X-ray observations of accreting supergiant X-ray binaries was already proven to be challenging, our study shows that similar imprints of clumps are found in the X-ray observations of the supergiant fast X-ray transients and at least one classical system, i.e. IGR J18027-2016. This provides interesting perspectives to further extend this study to many XMM-Newton observations already performed in the direction of other classical supergiant X-ray binaries.

\vspace{1cm}
\end{abstract}


\section{Introduction}
\label{sec: intro}

Supergiant high mass X-ray binaries (SgXBs) comprise a massive OB supergiant (typically $>$ 10 M$_{\odot}$) and, in most of the cases, a neutron star (NS) accreting from the companion's stellar wind. This wind can reach terminal velocities of several $\sim$1000 km s$^{-1}$ and  the mass loss rate from the supergiant star can be as high as 10$^{-4}$-10$^{-5}$~$M_{\odot}$~yr$^{-1}$ \citep[see, e.g.][for a recent review]{nunez17}. As the NSs in SgXBs are relatively young (a few 10$^{6}$~yrs), their magnetic field usually achieves strengths of 10$^{12}$~G, but much larger strengths cannot be ruled out \citep[see, e.g.][]{bozzo2008,rev15}. Due to this strong magnetic field, accreting material is channelled towards the poles of the NS and X-ray pulsations are observed as a consequence of the misalignment between the NS magnetic and rotation axes. Typical pulse periods span from a few to few thousands of seconds, while orbital periods of SgXBs have been measured in the range of a few to few tens of days \citep[see, e.g.,][]{walter2015}. 

An intriguing characteristic of X-ray emissions from SgXBs is the fast variability on time scales of few hundreds to few thousands seconds. This variability is usually ascribed to the properties of the stellar wind, which is well known to be largely in-homogeneous and populated by sub- and over-dense regions (``clumps'') created by an intrinsic instability of radiatively driven winds \citep[see, e.g.][]{puls08}. The suggestion that clumps could be the root cause of the pronounced variability of the SgXBs was proposed by \citet{sako2003}, who also suggested to use the NSs in these systems as {\em in-situ} probes of the stellar wind physical properties. In a direct wind-accreting system, any change in the density and/or velocity of the stellar wind is expected to produce a proportional change in the X-ray luminosity\footnote{We are not considering here and discussing in details the complications that might arise in certain conditions due to the presence of the NS rotating magnetosphere and/or the emergence of a settling accretion regime \citep{bozzo2008, shakura2012}.} which in turns affect the stellar wind through ionization and might led to further variations in its density and velocity \citep{kri15, kri16}. From an observational point of view, a clump simply passing in front of the NS (without being accreted) causes source dimming or even obscuration. Its presence can thus be revealed by the spectral signature of photoelectric absorption. Clumps that instead are intercepted by the NS lead to a temporarily larger mass accretion rate, giving rise to  
an X-ray outburst characterized by an enhanced local absorption that is proportional to the size of the accreted structure \citep[see, e.g.,][]{walter2007, intzand2005_sfxts, negueruela2008, bozzo13}. Probing the clump properties with NSs in SgXBs has recently received an increased attention due to the fact that constraining the parameters of these structures from the study of isolated supergiants has been proved challenging and clumps are known to largely affect the observationally derived mass loss rate of massive stars \citep[which in turns has implications in many fields of Astrophysics and Cosmology; see, e.g., discussions in][]{nunez17}. 

A substantial diversity in the physical properties of the clumps has been suggested to be one of the possible causes of the peculiar behavior in the X-ray domain of the so-called Supergiant Fast X-ray transients \citep[SFXTs; see][for a recent review]{walter2015}. These sources are a sub-class of SgXBs showing an enhanced variability in X-rays that can achieve a dynamic range of $\sim$10$^6$, i.e. a factor of 10$^3$-10$^4$ larger than the typical values measured in the classical SgXBs. The SFXTs spend most of their time in quiescence or in a low state, with typical X-ray luminosities as low as 10$^{32}$-10$^{33}$~\lerg, and only sporadically undergo bright X-ray outbursts lasting a few hours and reaching up to 10$^{38}$~\lerg \citep{romano2015_17544, bozzo2015, bozzo2016, sidoli2019}.   

Motivated by a number of source dimming events reported for SFXTs \citep[see, e.g.,][]{rampy2009,drave2014}, \citet{bozzo2017} carried out a systematic investigation of the physical properties of clumps in these sources by using a hardness-resolved spectral analysis of all available \xmm\ data where X-ray flares and outbursts were detected. The advantage of \xmm\ is that its unique combination of large effective area in the soft X-ray domain (0.5-10~keV) and good energy resolution allows us to obtain reasonably good spectra to measure statistically independent variations of the local absorption and spectral slope within time scales as short as few hundreds of seconds, i.e. following the rise and decay of most flares/outbursts. Furthermore, the long uninterrupted observations allow us to clearly distinguish among different flares/outbursts and periods of quiescence. Although most of these brightening events showed the imprints of clumps, the low number of recorded flares/outbursts and the generally low luminosity of the SFXTs prevented an accurate determination of the clump physical properties, and a monitoring program of all these sources with \xmm\ is presently on-going to increase the statistics of rapid spectral variations during flares/outbursts (Bozzo et al. 2019, in preparation). 

In this paper, we extend the study of \citet{bozzo2017} to the case of classical SgXBs in order to verify if also for these systems flares and outbursts in their lightcurves can be associated to the presence of clumps. While a number of previous studies in the literature have investigated sparse data from different facilities to indirectly search for clumps \citep[see, e.g.,][]{Kreykenbohm2008, ducci2009, fuerst2010, ducci2013, velax1_furst2014, pradhan2014}, a systematic study as the one carried out for the SFXTs is still missing. We use here the \xmm\ observations of the source \src\ as a test bench before conducting a  more enlarged systematic investigation of all other classical SgXBs. The \xmm\ data-set of \src\ is particularly interesting because we have five observations available covering different orbital phases within a single orbital revolution and an additional isolated observation performed few years later. Although these data have been already analysed \citep{walter2006}, we report on them here for completeness. The latter is also re-analyzed here for completeness. We report a short description of \src\ in Sect.~\ref{sec: src}, and describe our \xmm\ data analysis procedure and results in Sect.~\ref{sec: data}. Our discussion and conclusions are reported in Sect.~\ref{sec: results} and \ref{sec: disc}, respectively. 

\section{The classical SgXB \src\ }
\label{sec: src}

The classical SgXB \src (alias SAX J1802.7--2017) was discovered with the IBIS/ISGRI instrument on-board \inte\ during the Galactic Center survey \citep{rev2004}. The distance to the source is estimated to be 12.4 kpc \citep{torrejon2010_dist}. The system hosts a NS spinning at $\sim$139~s \citep{iaria2004R} and orbiting around a B1-Ib supergiant companion \citep{torrejon2010_dist}. The measured orbital period of the system is $\sim$4.57~d  \citep{hill2005,jain2009}. This source was observed several times with different X-ray facilities, and its broad-band X-ray spectrum can be generally characterized by a broken power law modified by a photoelectric absorption with a typical hydrogen column density of N$_{H}$ $\sim$10$^{23}$~cm$^{-2}$ \citep{hill2005}. The presence of a possible soft excess below 2~keV has also been reported different times \citep{hill2005, walter2006}. The long-term evolution of the orbital period of the source has been studied in detail by \citet{falanga2015} and \citet{coley2015} but no significant orbital period derivative has been reported. 

A relatively recent monitoring campaign with \swift\ /XRT covering multiple orbits at sparse orbital phases have shown large X-ray intensity variations on timescales of hundreds to thousands seconds (as expected for a wind-accreting classical SgXB), as well as significant spectral variations (in the powerlaw slope and the absorption column density) at different orbital phases \citep{aftab2016}.

\section{Data analysis and results}
\label{sec: data}

\xmm\ observed \src\ six times, including an earlier isolated exploratory observation performed after the discovery of the source (OBSID~0206380601) and five more recent observations probing different epochs of the same orbital revolution (OBSID~0745060401, 0745060501, 0745060601, 0745060701, 
0745060801). Out of these last five observations, one caught the source during the mid and egress from the eclipse (OBSID~0745060401), while one observation was carried out during the entrance to the eclipse (OBSID~0745060701). Following \citet{hill2005} and considering that no significant orbital period evolution has been reported so far, we fixed our reference mid-eclipse time at 52931.37~MJD (phase 0 in the following). The coverage of source orbital phases of all observations is reported in Table~\ref{tab: log}, which also provides a log of all used observations.
\begin{table}[ht!]
\flushleft
 \caption{Log of all \xmm\ observations of \src\ used in the current paper along with their exposure. Orbital phases were determined by folding the pn light curves assuming an orbital period of P$_{orb}$=394851.65~s (as determined from the \swift/BAT long-term lightcurve of the source, see Fig.~\ref{bat_xmm}) and a phase 0 centered on the mid-eclipse time 52931.37~MJD \citep{hill2005}. We also tabulate the spin periods obtained from the individual \xmm\ observations.}
\scriptsize
\begin{tabular}{|c|c|c|c|c|c|}
\hline
OBSID & START TIME & Exp. & $\phi$ & P$_{spin}$ \\
      &  (MJD) & (ks) & & (s) \\
\hline
0206380601 & 53101.2738 & 11.9 & 0.172-0.203 & 139.6 $\pm$ 0.4\\
0745060401 & 56911.9380 & 46.0 & 0.000-0.125 & 139.8\footnote{only during egress} $\pm$ 0.3 \\
0745060501 & 56908.8321 & 19.0 & 0.328-0.375 & 139.8 $\pm$ 0.3 \\
0745060601 & 56909.9583 & 20.0 & 0.578-0.625 & 140.0 $\pm$ 0.3\\
0745060701 & 56906.4133 & 17.3 & 0.781-0.843 & 140.04 $\pm$ 0.4 \\
0745060801 & 56912.8112 & 19.5 & 0.219-0.250 & 139.8 $\pm$ 0.3\\
\hline
\end{tabular}
\label{tab: log}
\end{table}  

\xmm\ observation data files (ODFs) were processed by using the standard Science Analysis System (SAS 17.0) and following the procedures given in the on-line analysis threads\footnote{\scriptsize{http://www.cosmos.esa.int/web/xmm-newton/sas-threads}}. 
We used data from the EPIC-pn, as well as EPIC-MOS, in all but the observation ID.~0206380601 where the source laid outside the MOS CCDs and only pn data were available. We did not make use of the RGS data, due to the limited band-pass of this instrument and the heavy X-ray extinction in the direction and local to the source (see Sect.~\ref{sec: results}). We checked for pile-up from all the individual observations using the SAS tool {\sc epaplot} and did not find any significant issue. We also verified the presence of high flaring background time intervals and found that only the OBSID~0206380601 was significantly affected, as also previously reported by \citet{walter2006}. We thus excluded the affected high background time intervals from this observation before carrying out any other analysis (we used the standard cut in count-rate suggested in the online SAS data analysis threads). The resulting effective exposure time was of 5.6~ks. For all other observations, we retained the entire exposure time available and extracted first the barycentered corrected event files of the source and background to detect and measure the source spin period. 

\begin{figure*}[ht!]
 \centering
 \includegraphics[width=6.5cm,angle=0]{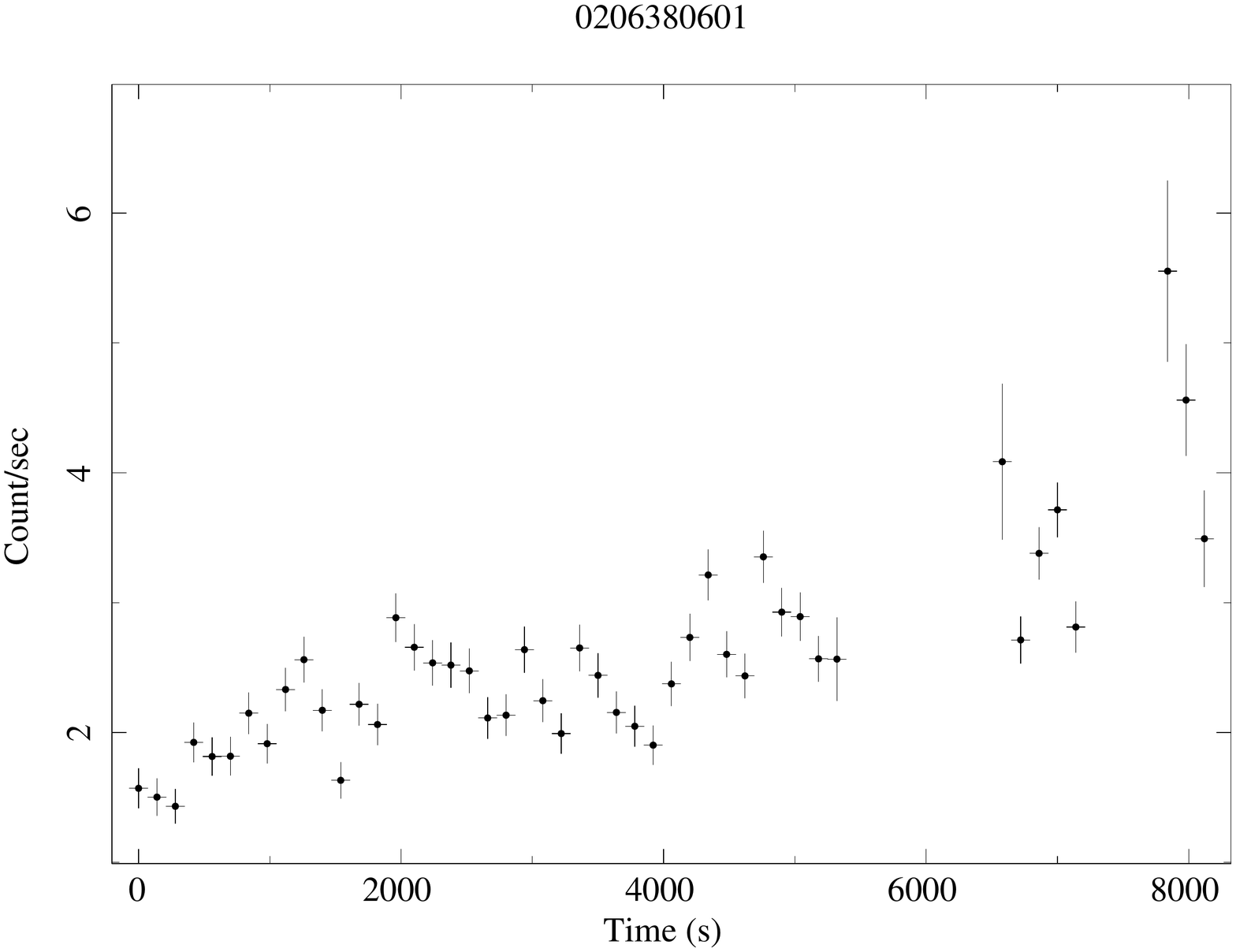}
 \includegraphics[width=6.3cm,angle=0]{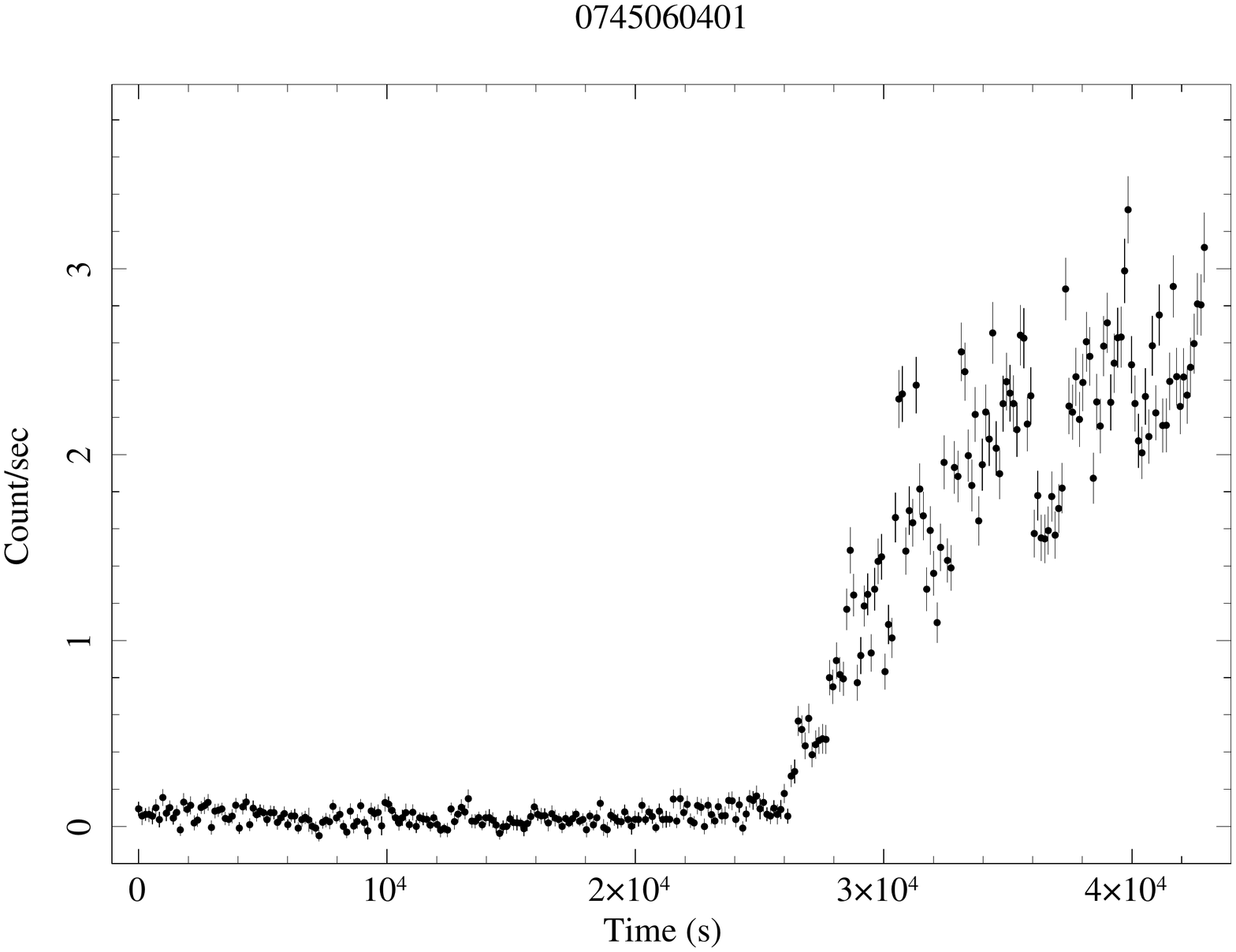}
 \includegraphics[width=6.5cm,angle=0]{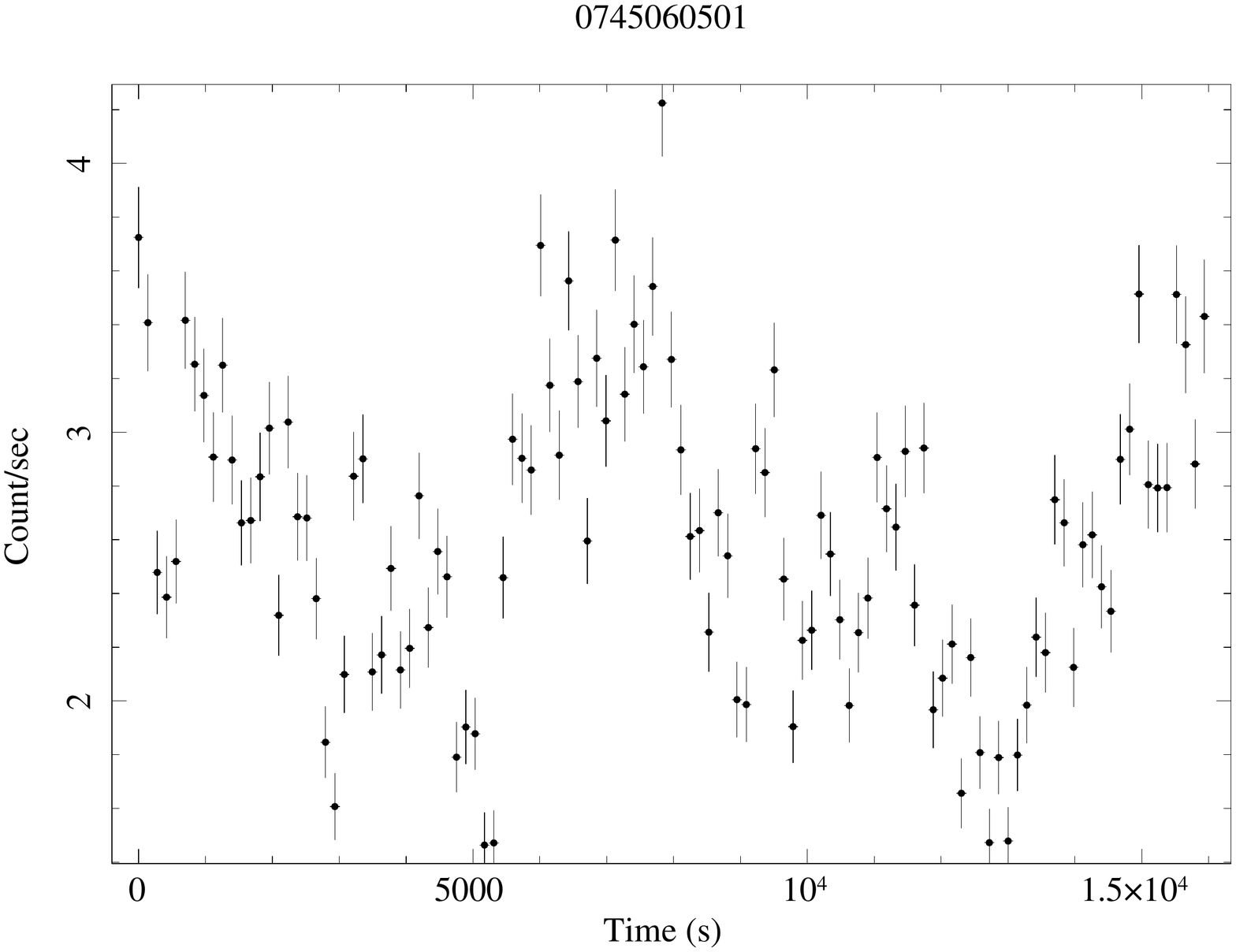}
 \includegraphics[width=6.5cm,angle=0]{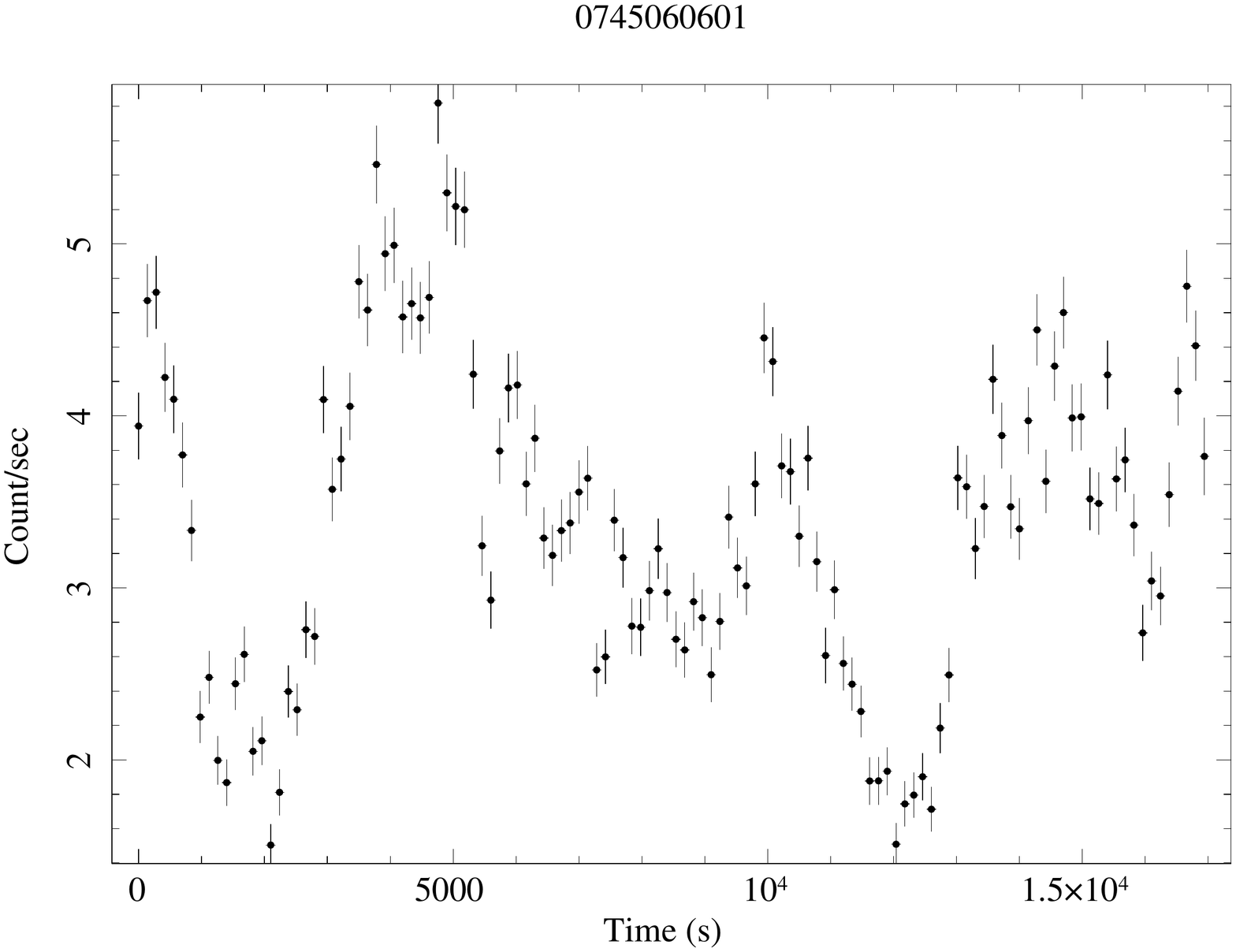}
 \includegraphics[width=6.7cm,angle=0]{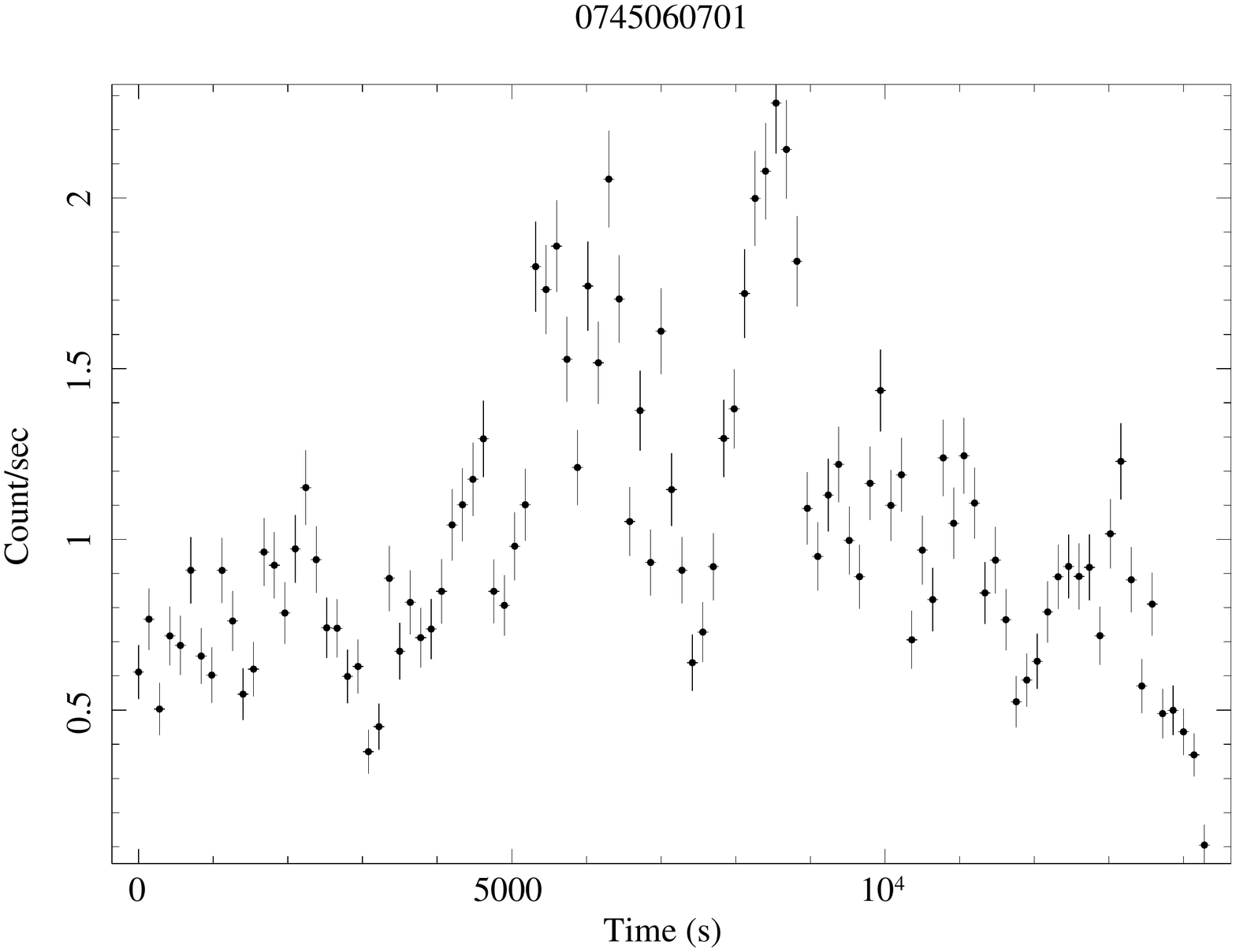}
 \includegraphics[width=6.7cm,angle=0]{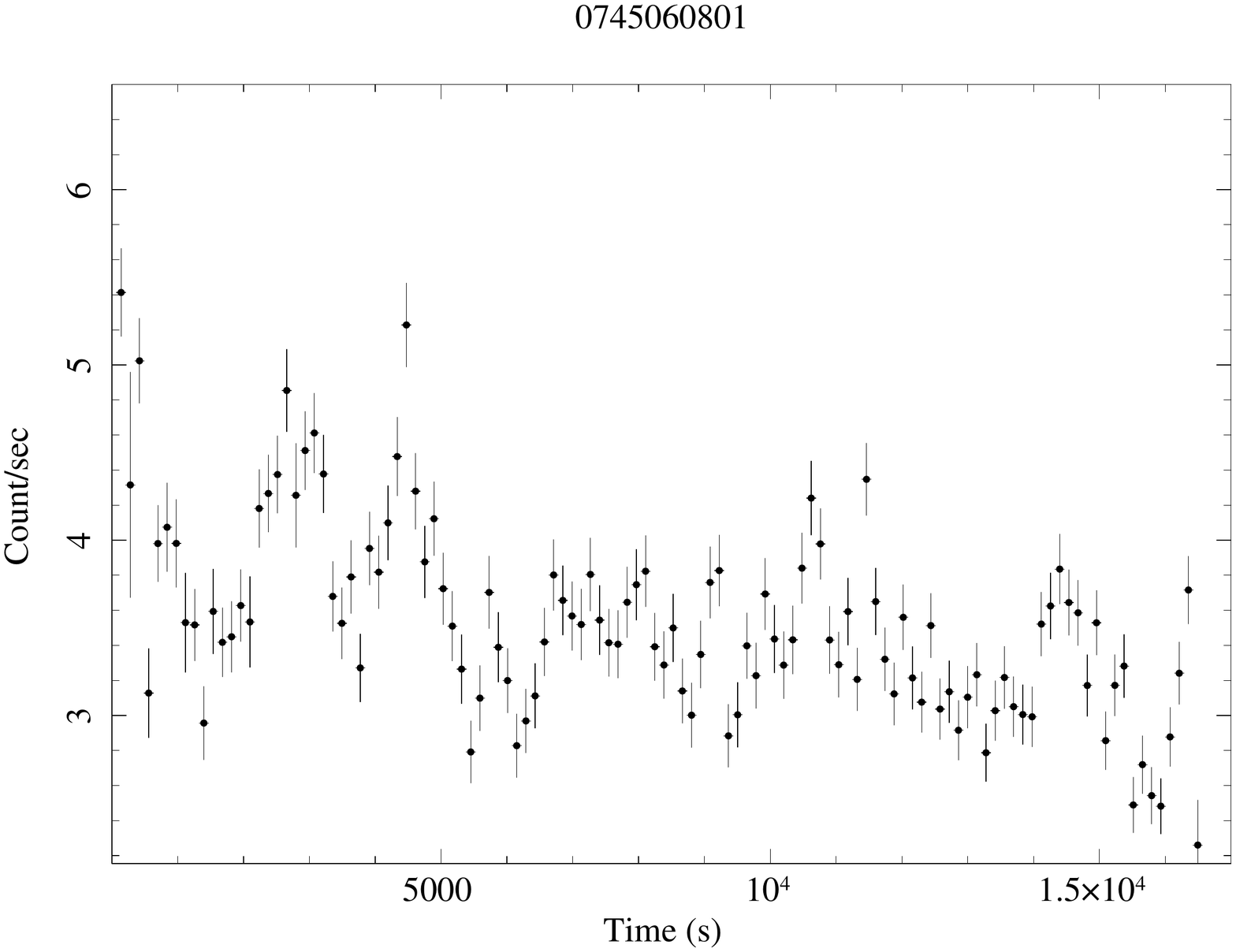}
 \label{lcurve}
 \caption{Lightcurves of \src\ as observed by the \xmm\ EPIC-pn camera (0.5-10~keV) during the different observations used in this paper. The time bin in each case is the same as the best determined pulse period of the corresponding observation (see text for details).}
\end{figure*}

The limited length of observations and the relative slow pulsation of the source prevented us from determining the spin period with a precision better than a fraction of a second. For typical masses of the supergiant stars and the measured orbital period, the spin period variations due to orbital Doppler shift are of the order of the expected period uncertainty. For the sake of this work, we determined the spin periods using and epoch folding technique \citep{fold}.  
We obtained a spin period of 139.6$\pm$0.4~s from the OBSID~0206380601, 139.8$\pm$0.3~s from the OBSID~0745060501, 140.0$\pm$0.3~s from the OBSID~0745060601, 140.04$\pm$0.4~s from the OBSID~0745060701, and 139.8$\pm$0.3~s from the OBSID~0745060801. In the OBSID~0745060401, pulsations were not detected during the first 26~ks when the source was still in eclipse. However, pulsations could be significantly detected using the last 18~ks of data when the source was caught during the egress from the eclipse. In this case, we measured with the same technique above a spin period of 139.8$\pm$0.3~s. These spin periods have also been reported in Table \ref{tab: log}.
As in this paper we are not focusing on any spin resolved spectral analysis because we want to study possible spectral variations on longer time scales related to the presence of clumps, we binned the pn 0.5-10~keV lightcurve of each observation at the correspondingly best determined spin period to level out the variability caused by the NS rotation.
These lightcurves, shown in Fig.~\ref{lcurve}, evidence a remarkable X-ray variability on time scales of several hundreds to thousands seconds, typical of wind accreting systems and classical SgXBs (see Sect.~\ref{sec: intro}). For completeness, we report the energy-resolved pulse profiles of the source in Appendix~\ref{sec:appendix}.  

\begin{figure}
\centering
\includegraphics[width=8cm,angle=0]{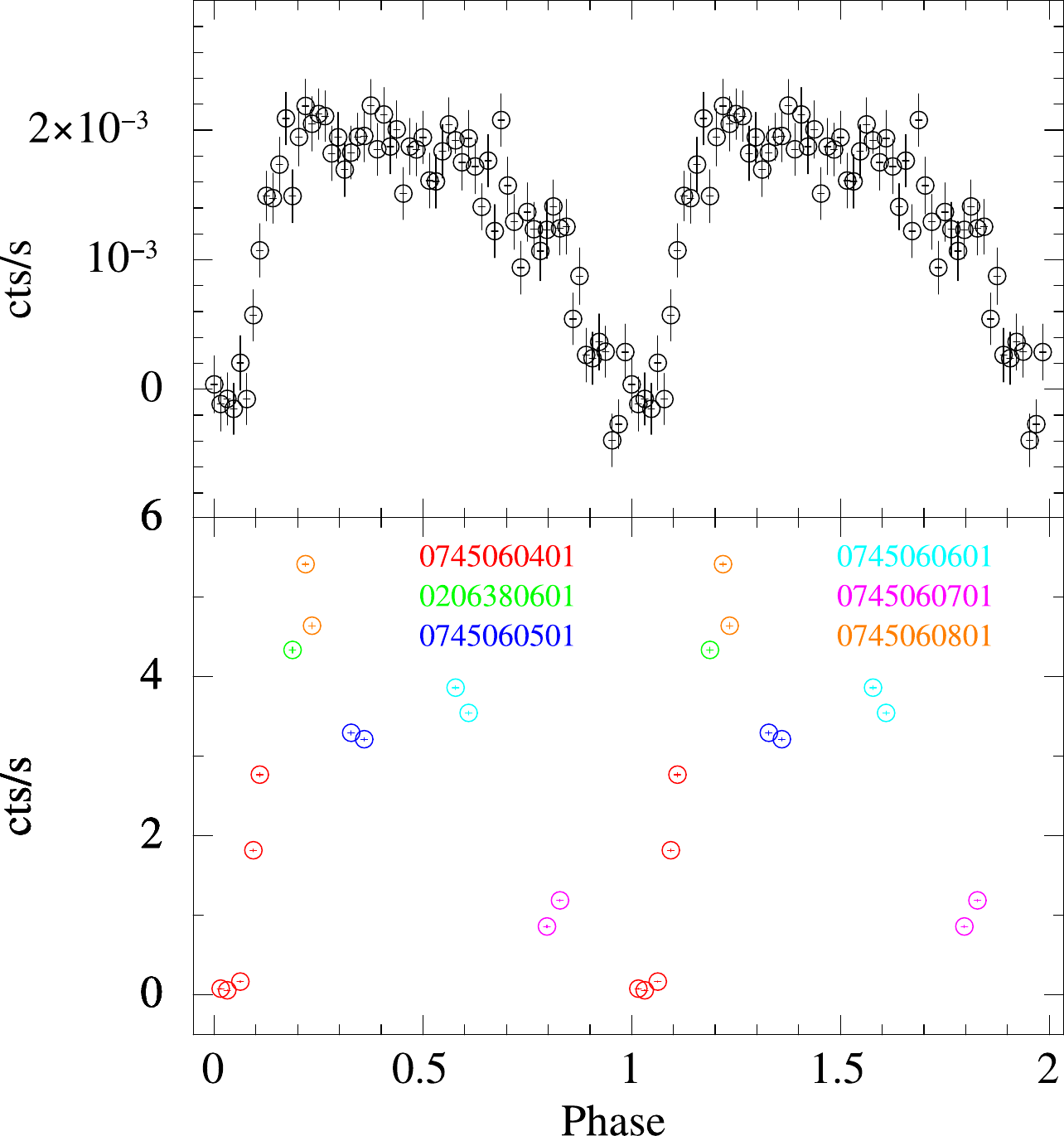}
\caption{\swift/BAT and \xmm\ lightcurves of \src\ folded at the best determined orbital period measured from the BAT data 
(P$_{orb}$=394851.65\,s) and assuming phase 0 at the mid-eclipse time 52931.37~MJD \citep{hill2005} }
\label{bat_xmm}
\end{figure}

An average spectrum for each observation was extracted from the pn and MOS data to measure the emission properties of the source in the \xmm\ energy band. The source spectra were extracted from a region centered around the best known source position \citep{chaty08} and with a variable radial extension between 600 and 1000 pixels, depending on the source intensity. Background spectra were extracted from a region located on the same CCD as that used for the source but carefully checking to minimize the contamination from the source emission. The difference in extraction areas used for the source and background was accounted for by using the SAS {\sc backscale} task. All spectra were rebinned in order to have at least 25 counts per energy bin and, at the same time, to prevent oversampling of the energy resolution by more than a factor of three. In case of the OBSID~0745060401, we extracted two spectra: one during the eclipse period (the first 26~ks) and one during the egress (the last 18~ks). The pn and MOS data from each observation were fit simultaneously (we used \texttt{XSPEC} v12.10.0e) with a power-law model corrected for the line of sight Galactic and local absorption with a \texttt{phabs} component, as typically done in case of classical SgXBs \citep{walter2015}. The addition of a partial covering (\texttt{pcfabs}), usually required in these sources as due to part of the X-ray emission from the NS that escapes the local absorption and it is affected only by the Galactic extinction, was required to account for the soft excess and achieve an acceptable fit for the higher statistics spectra (OBSID~0745060401 - egress, 0745060501, 0745060601, and 0745060701). Two Gaussian-shaped emission features were also added to the best fit models of most spectra and are compatible with being the iron K$\alpha$ and K$\beta$ lines. These are known to be present in the X-ray spectra of classical SgXBs and are produced by the fluorescence of the X-rays from the accreting NS onto the surrounding stellar wind material. In the case of the OBSID~0745060701, we also found the presence of a second K$\alpha$ lines corresponding to an ionized state of iron. This is characterized by a centroid energy of $\sim$6.7~keV and it is likely associated to Fe XXV \citep[this feature is not uncommon in wind-fed high mass X-ray binaries; see, e.g.,][]{torrejon}. Note that for the absorption models \texttt{phabs} and \texttt{pcfabs}, we used the default element abundances and cross-sections in \texttt{XSPEC} \citep[]{anders1989,verner1996}. We verified in all cases that the usage of different absorption models \citep[e.g., {\sc TBabs}]{wilms00} and/or different abundances and cross-sections did not significantly affect the results, given the relatively low S/N of all data at energies $\lesssim$2-3~keV. The spectra from all observations and the corresponding best fits, including the residuals from these fits, are shown in Fig.~\ref{averagespectra}. All results are given in Table~\ref{spec}. 
We note that for all those observations requiring a partial absorption component, an equivalently good fit could be obtained by using a thermal black-body component with a temperature of $\sim$0.2~keV and a radius of $\sim$100~km to describe the soft-excess. We did not include the results from these fits in Table~\ref{spec} because the measured radius would be more compatible with that of an accretion disk around the NS rather than an hot spot on the compact object, as expected for a wind-fed system like \src.\ 
\begin{table*}[t!]
\begin{center}
 \caption{Best fit spectral parameters measured from the source averaged spectrum in each of the analyzed observations. Uncertainties are quoted at 90 \% confidence (as everywhere else in the paper, unless stated otherwise). Orbital phases ($\phi$) are determined by folding the pn light curves with P$_{orb}$=394851.65\,s and assuming phase 0 at the mid-eclipse time 52931.37~MJD \citep{hill2005}. Note that the absorption column densities $N_{\rm H1}$ and $N_{\rm H2}$ are in units of $10^{22}$~$\rm{cm^{-2}}$, while the flux is given in the 1-10 keV energy band in units of $\times$ $10^{-11}$~\ferg (we also report the correspondingly calculated X-ray luminosity, $L_{\rm X}$, in units of $10^{35}$~erg~s$^{-1}$ assuming a distance of 12.4~kpc).}
\begin{tabular}{|c|c|c|c|c|c|c|c|}
\hline
Spectral Parameter & \multicolumn{7}{c}{OBSID} \\
\hline
& 0206380601 & \multicolumn{2}{c|}{0745060401} & 0745060501 & 0745060601 & 0745060701 & 0745060801 \\
\hline
& & (eclipse) & (egress) & & & & \\
\hline
$\phi$ & 0.172-0.203 & 0.000-0.0045 & 0.045-0.125 & 0.328-0.375 & 0.578-0.625 & 0.781-0.843 & 0.219-0.250 \\
$N_{\rm H1}$ & 9.69 $\pm$ 0.60 & $<$0.8 & 5.50 $\pm$ 0.90 & 2.34 $\pm$ 0.20 & 2.80 $\pm$ 0.70 & 2.53 $\pm$ 0.18 & 1.85 $\pm$ 0.04 \\
$N_{\rm H2}$ & - & - & 12.20 $\pm$ 2.90 & 6.02 $\pm$ 1.12 & 7.08 $\pm$ 0.46 & 53.79 $\pm$ 0.70 & - \\
$f$ & - & - & 0.70 $\pm$ 0.09  & 0.55 $\pm$ 0.05 & 0.88 $\pm$ 0.05 & 0.95 $\pm$ 0.01  & - \\
$\Gamma$ & 0.94 $\pm$ 0.09 & -0.2$\pm$0.3 & 0.84 $\pm$ 0.08 & 1.00 $\pm$ 0.05 & 0.82 $\pm$ 0.05 & 0.66 $\pm$ 0.04 & 0.99 $\pm$ 0.02 \\
K$\alpha$ (keV) & - & 6.50 $\pm$ 0.04 & 6.44 $\pm$ 0.03  & 6.38 $\pm$ 0.05 & 6.40 $\pm$ 0.01 & 6.43 $\pm$ 0.02 & 6.42 $\pm$ 0.03 \\
 &  & &   &  &  & (6.66 $\pm$ 0.06) &  \\
EW (keV) & - & 1.7 $\pm$ 0.3 & 0.03 $\pm$ 0.01 & 0.09 $\pm$ 0.01  & 0.10 $\pm$ 0.01 &  0.20 $\pm$ 0.03 & 0.06 $\pm$ 0.01 \\
 &  &  &  &  &  &  (0.06 $\pm$ 0.02) & - \\
K$\beta$ (keV) & - & - & - & 6.97 $\pm$ 0.07 & - & 7.03 $\pm$ 0.02 & - \\
EW (keV) & - & - & -  & 0.02 $\pm$ 0.01 & - & 0.09 $\pm$ 0.02 & - \\
Flux & 3.5 $\pm$ 0.1 & 0.082 $\pm$ 0.007 & 2.6 $\pm$ 0.1  & 3.5 $\pm$ 0.1 & 4.7 $\pm$ 0.1 & 1.8 $\pm$ 0.1 & 4.9 $\pm$ 0.1 \\
L${_X}$ & 6.1 $\pm$ 0.3  & 0.14 $\pm$ 0.01   & 4.5 $\pm$ 0.3 & 6.1 $\pm$ 0.3 & 8.2 $\pm$ 0.3 & 3.1 $\pm$ 0.2 & 8.5 $\pm$ 0.3 \\
$\chi^{2}_{red}$/d.o.f & 1.02/121 & 1.13/174 & 1.08/356 & 1.10/454 & 1.03/424 & 1.15/326 & 1.16/482 \\
\hline
\end{tabular}
\label{spec}
\end{center}
\end{table*}

\begin{figure}
\centering
\includegraphics[width=9.5cm,angle=0]{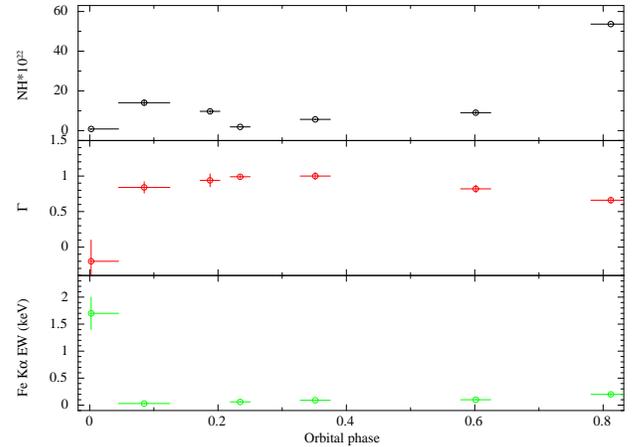}
\caption{Average \nh (=$N_{\rm H1}$+$f$$\times$$N_{\rm H2}$), $\Gamma$, and EW of the Fe K$_\alpha$ line for different orbital phases as reported in Table~\ref{spec}. The orbital phases have been calculated as in Fig.~\ref{bat_xmm}. The time-resolved variations of the spectral parameters for each \xmm\ observation are shown in Fig.~\ref{averagespectra}.}
\label{spec_xmm}
\end{figure}

\begin{figure*}[t!]
 \centering
 \includegraphics[width=7.5cm,angle=0]{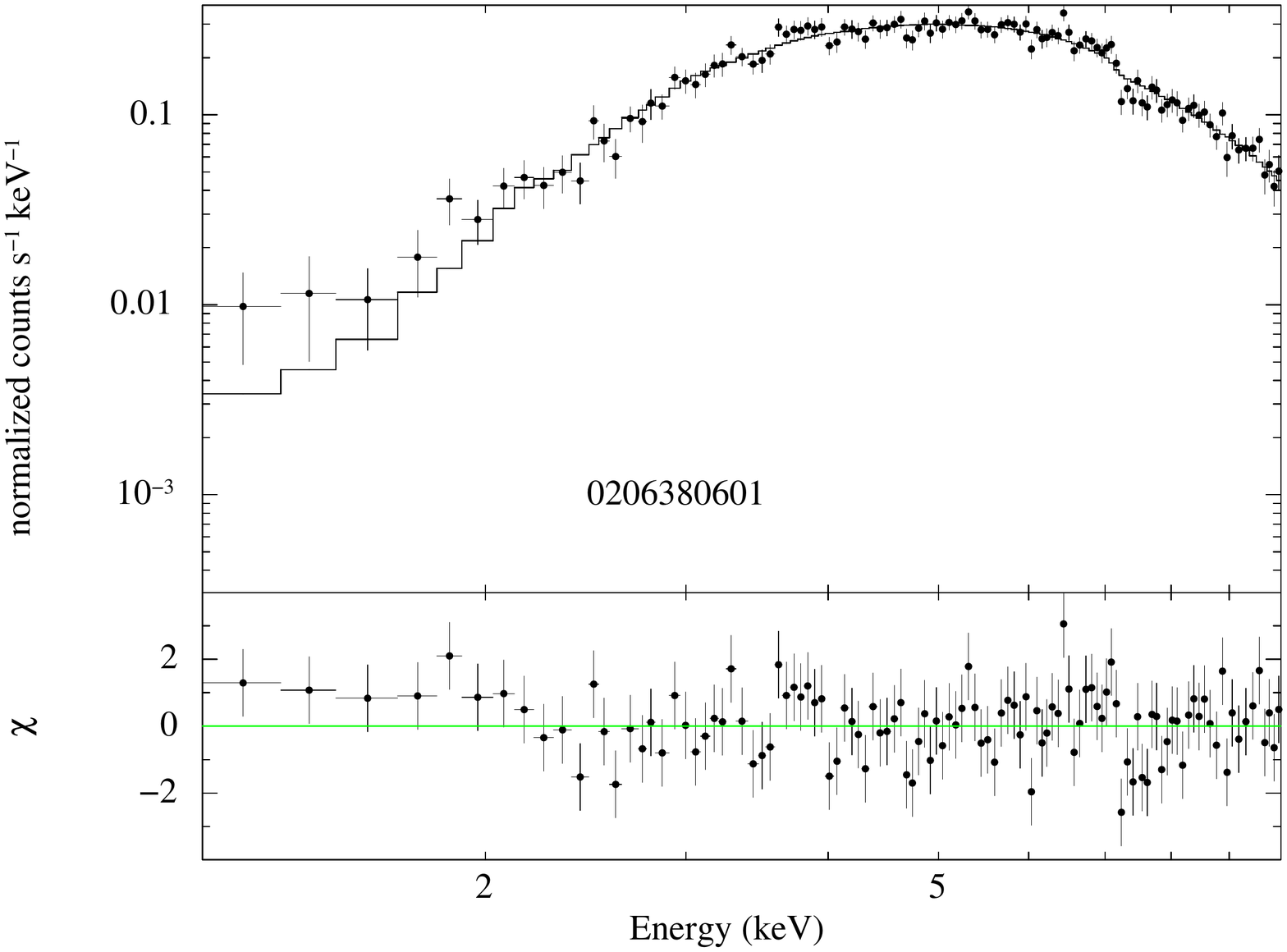}
 \includegraphics[width=7.5cm,angle=0]{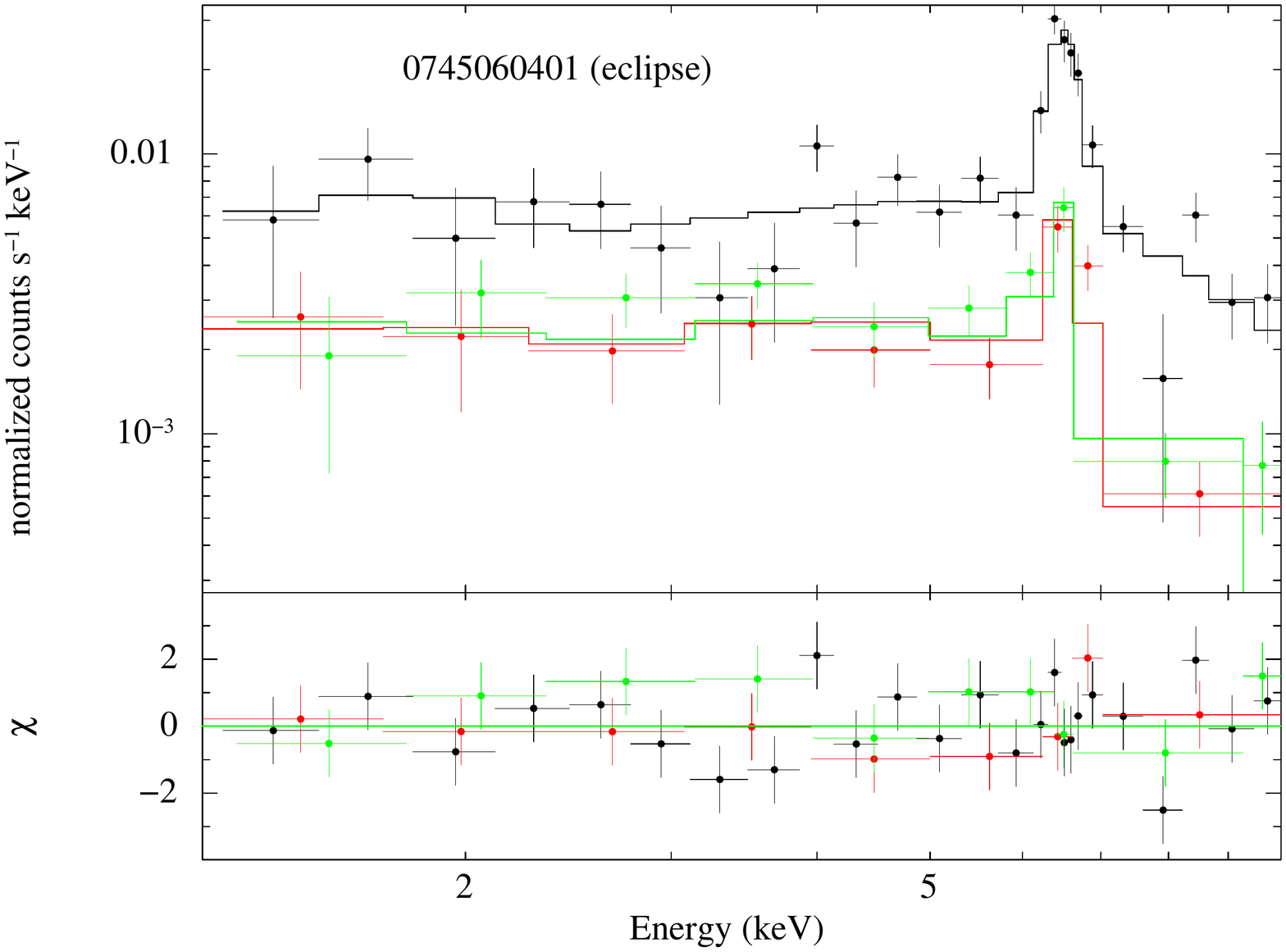}
 \includegraphics[width=7.5cm,angle=0]{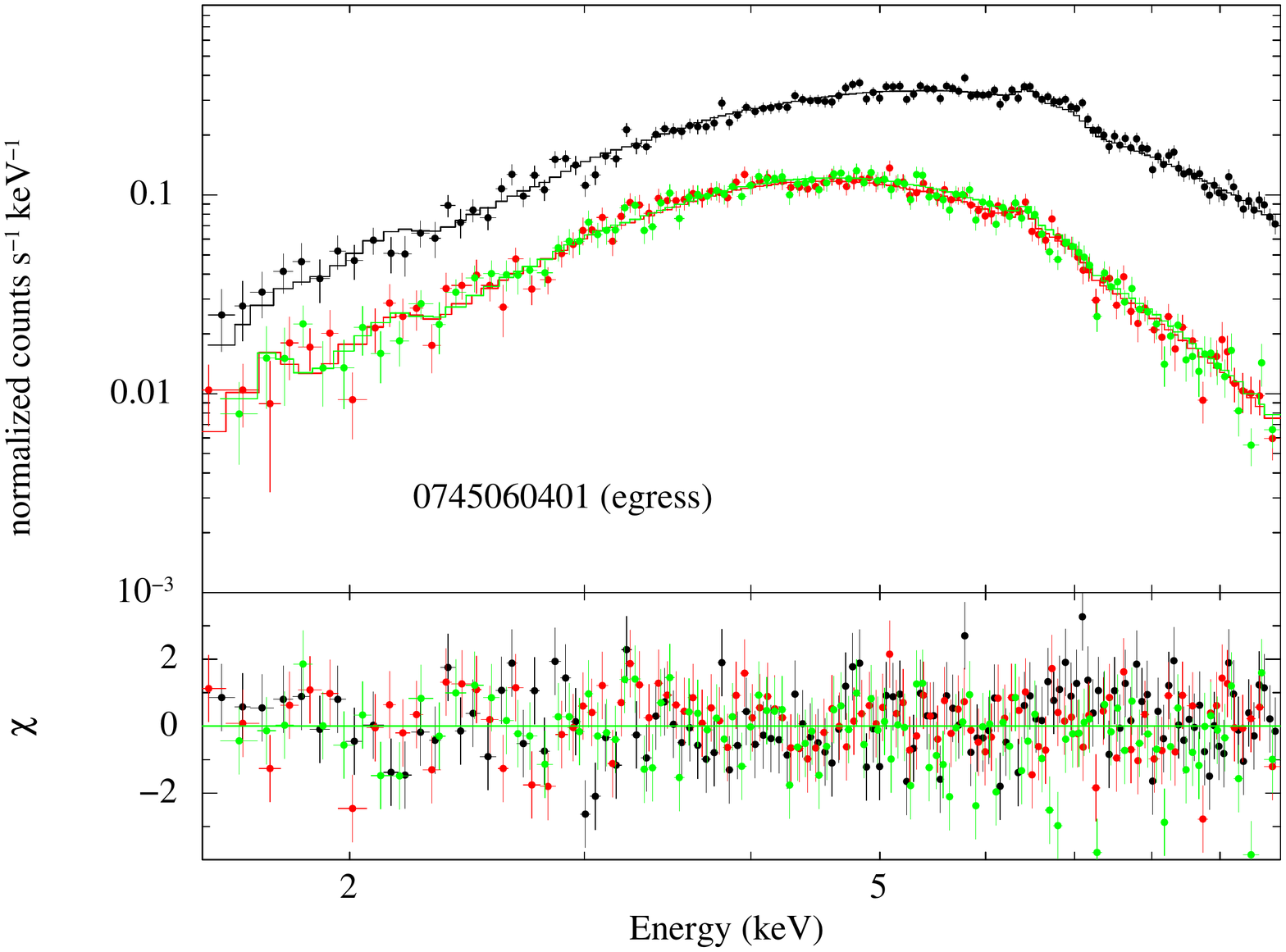}
 \includegraphics[width=7.5cm,angle=0]{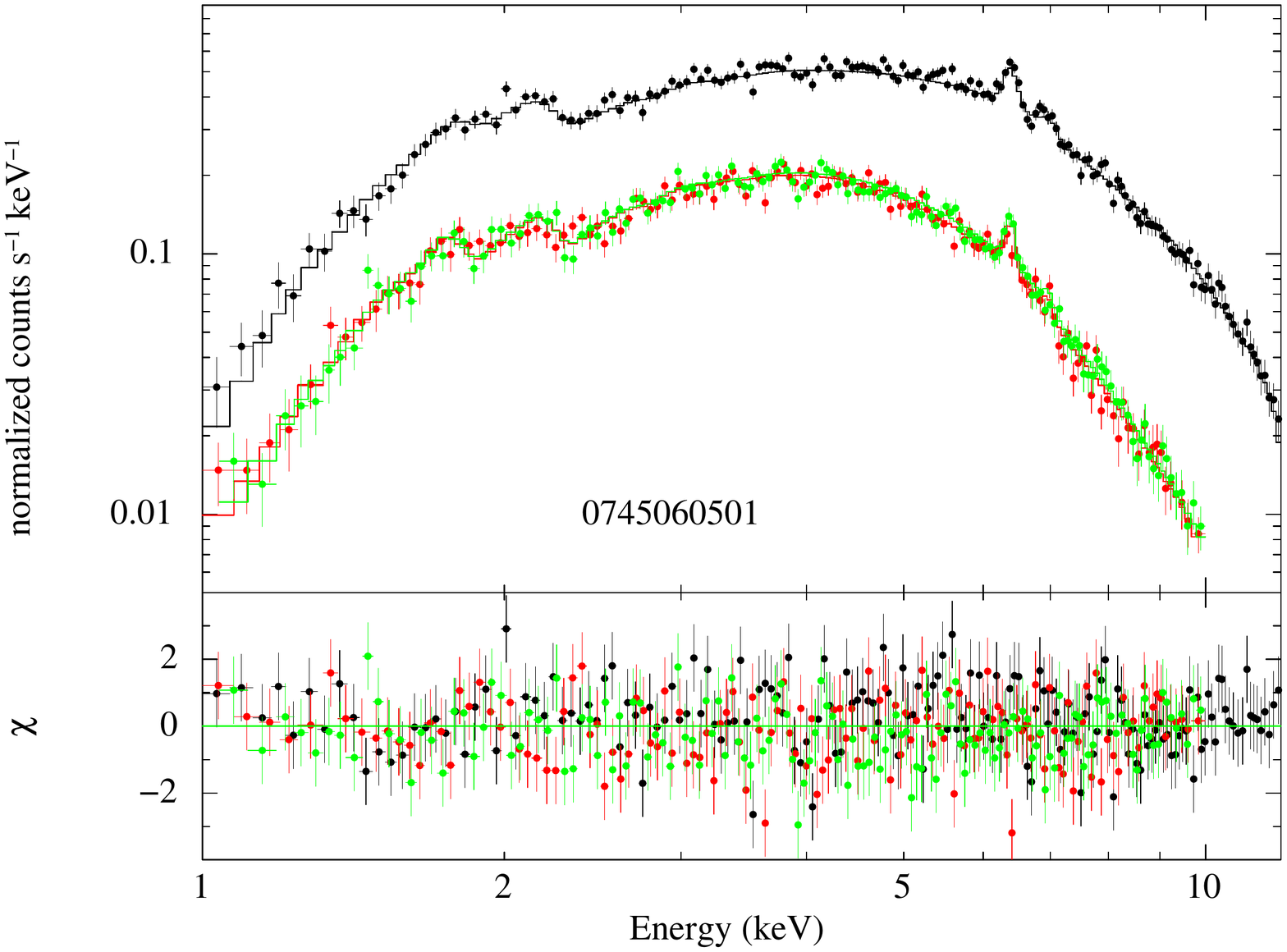}
 \includegraphics[width=7.5cm,angle=0]{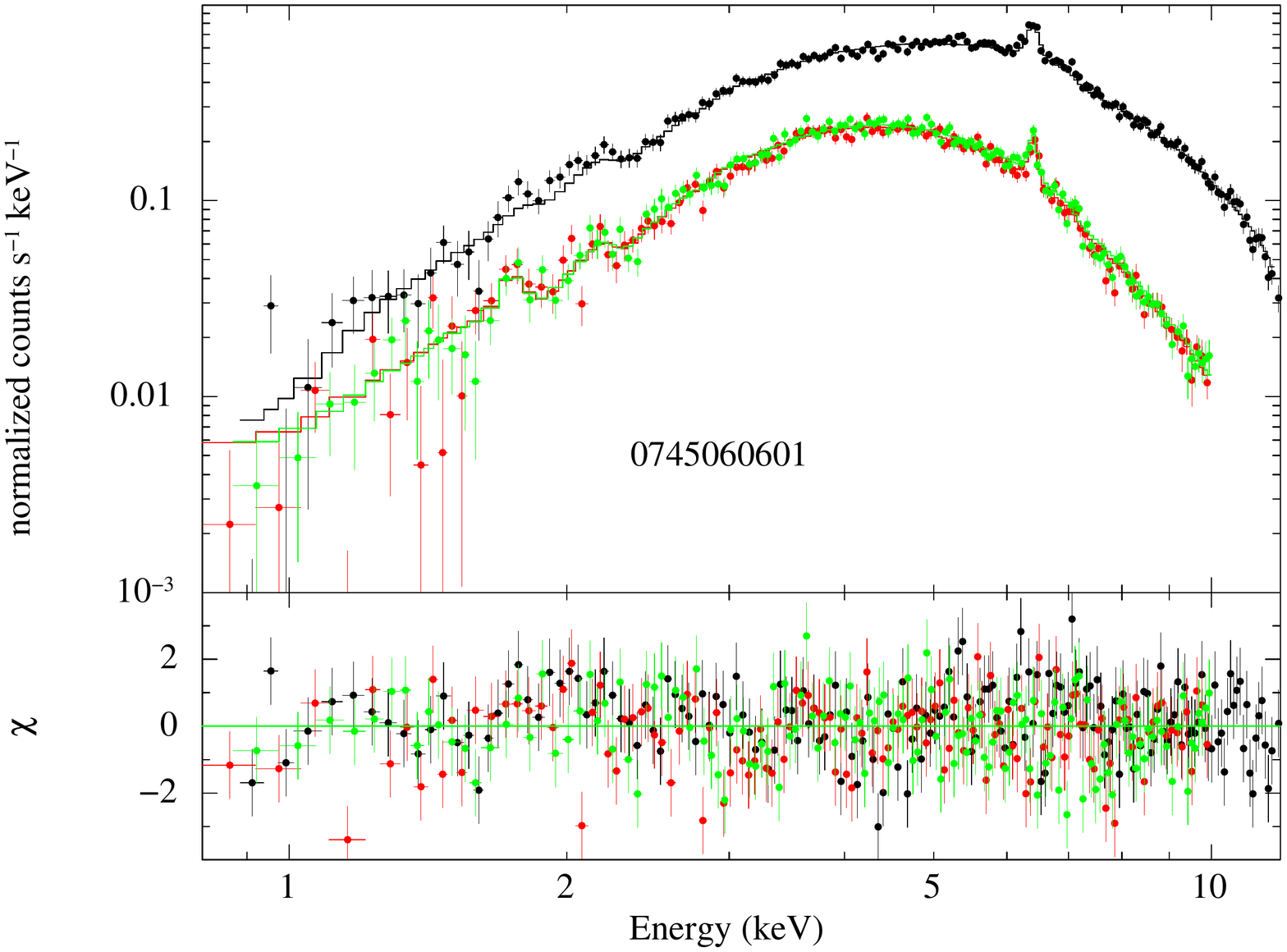}
 \includegraphics[width=7.5cm,angle=0]{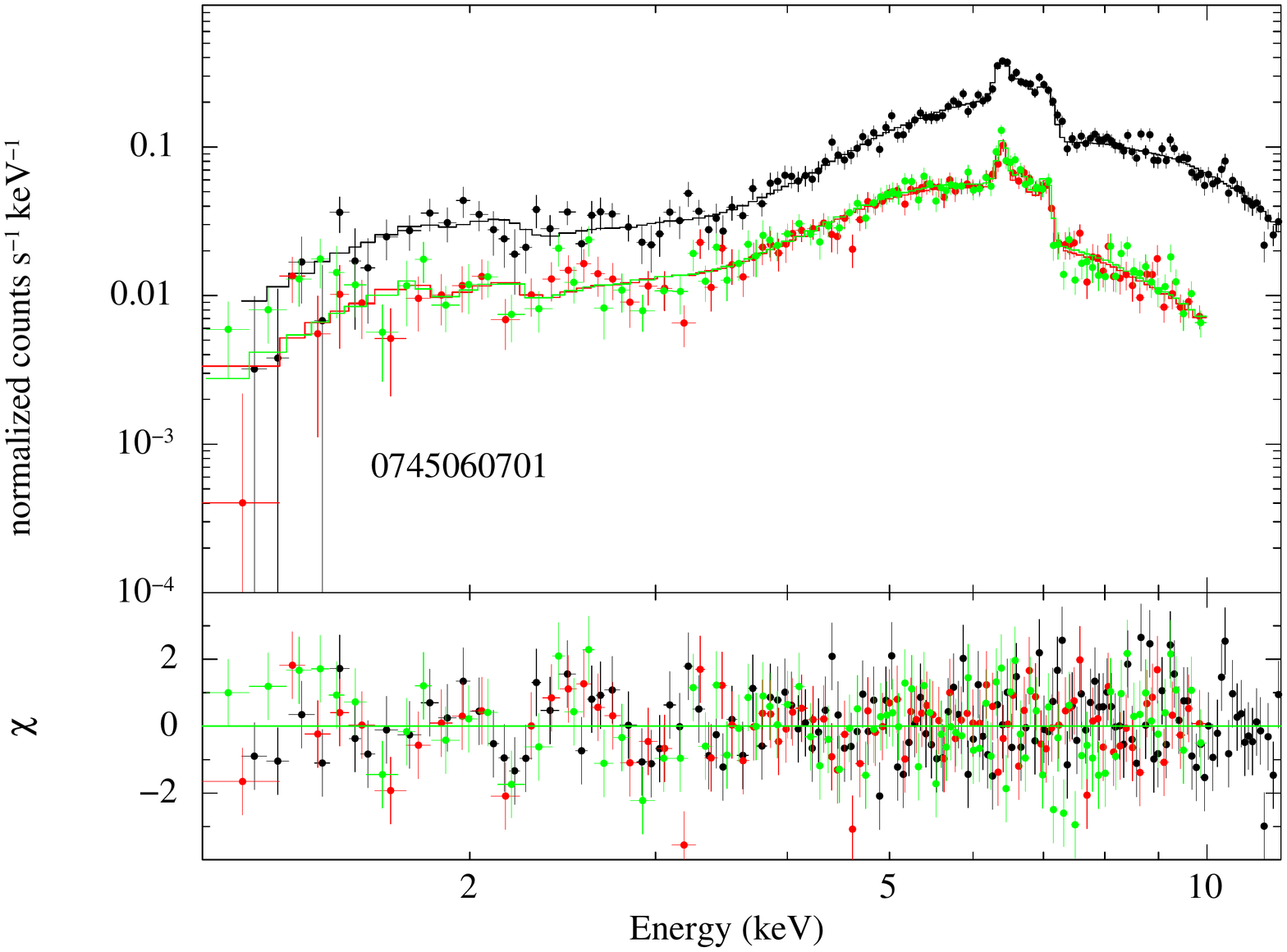}
 \includegraphics[width=7.5cm,angle=0]{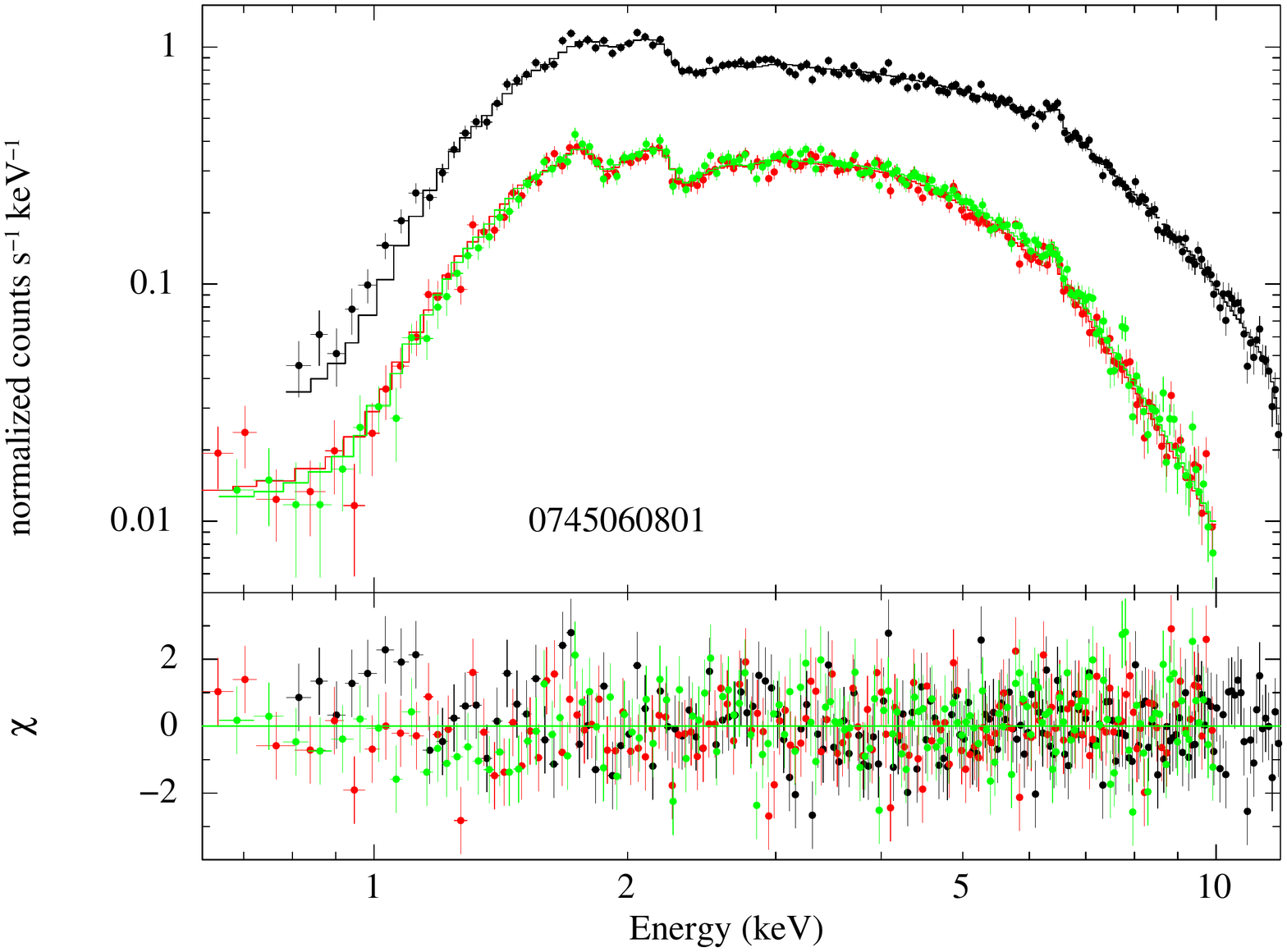} 
 \caption{Averaged spectra of \src\ extracted from each available \xmm\ observation by using pn (black), MOS1 (red), and MOS2 (green) data. A simultaneous fit of the three spectra has been performed for each observation and all results are reported in Table~\ref{spec}. The residuals from the best fits are shown in the bottom panel of each spectrum figure. In the case of the observation 0745060401, two spectra were extracted during the eclipse period and the egress from the eclipse (see also Fig.~\ref{lcurve}).}
 \label{averagespectra}
 \end{figure*}

\begin{figure*}
 \centering
 \includegraphics[width=6cm,height=7cm,angle=0]{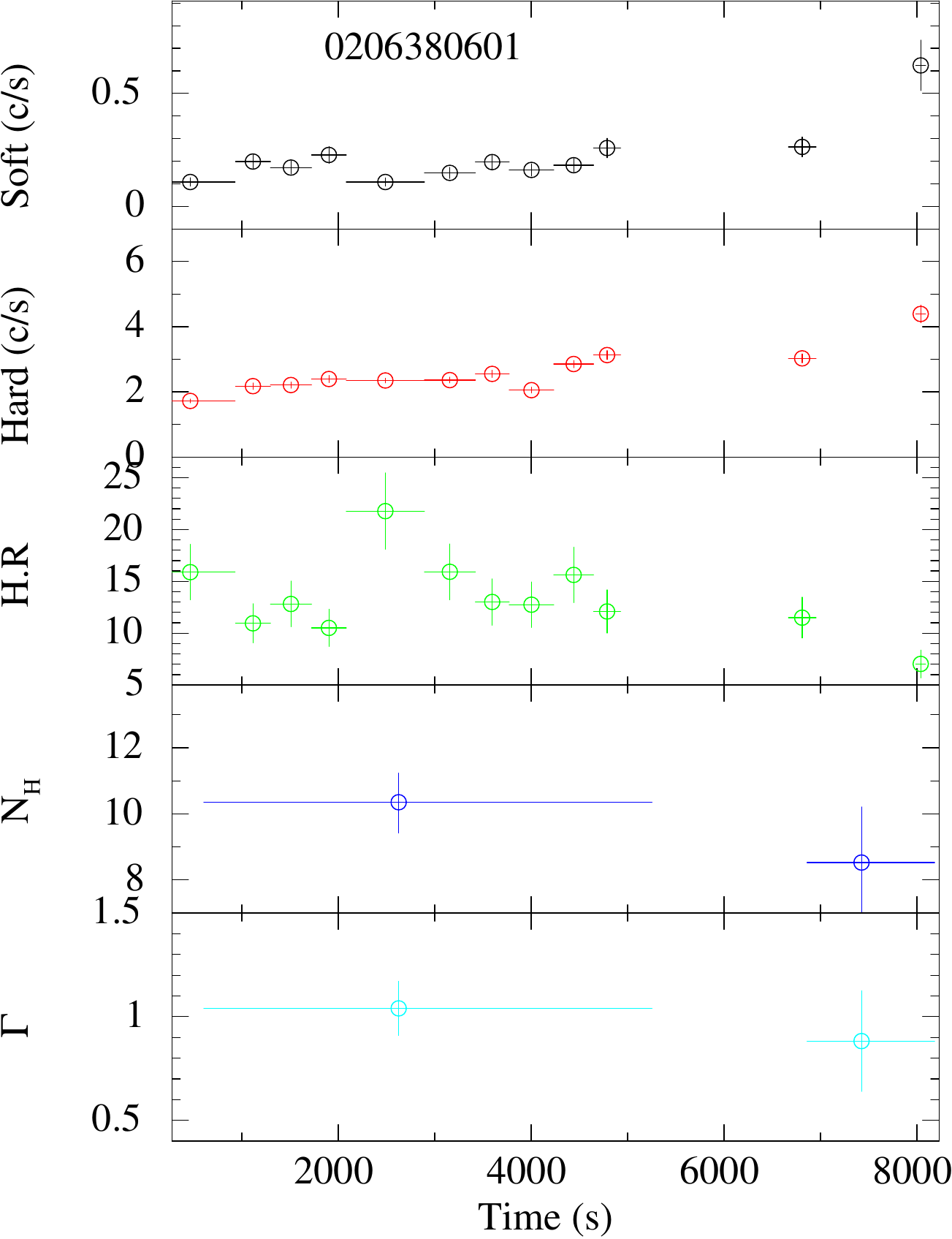}
 \includegraphics[width=6cm,height=7cm,angle=0]{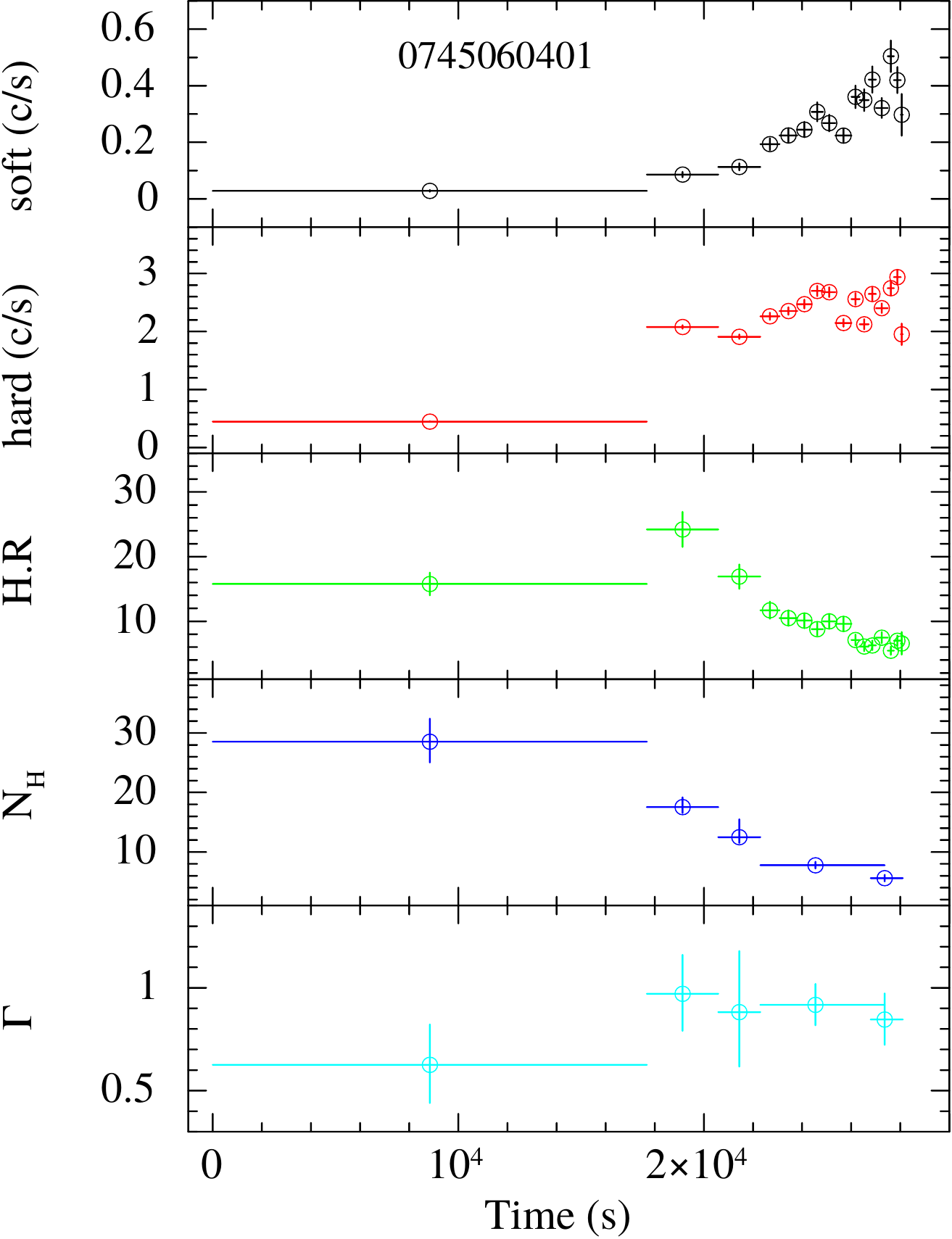}
\includegraphics[width=6cm,height=7cm,angle=0]{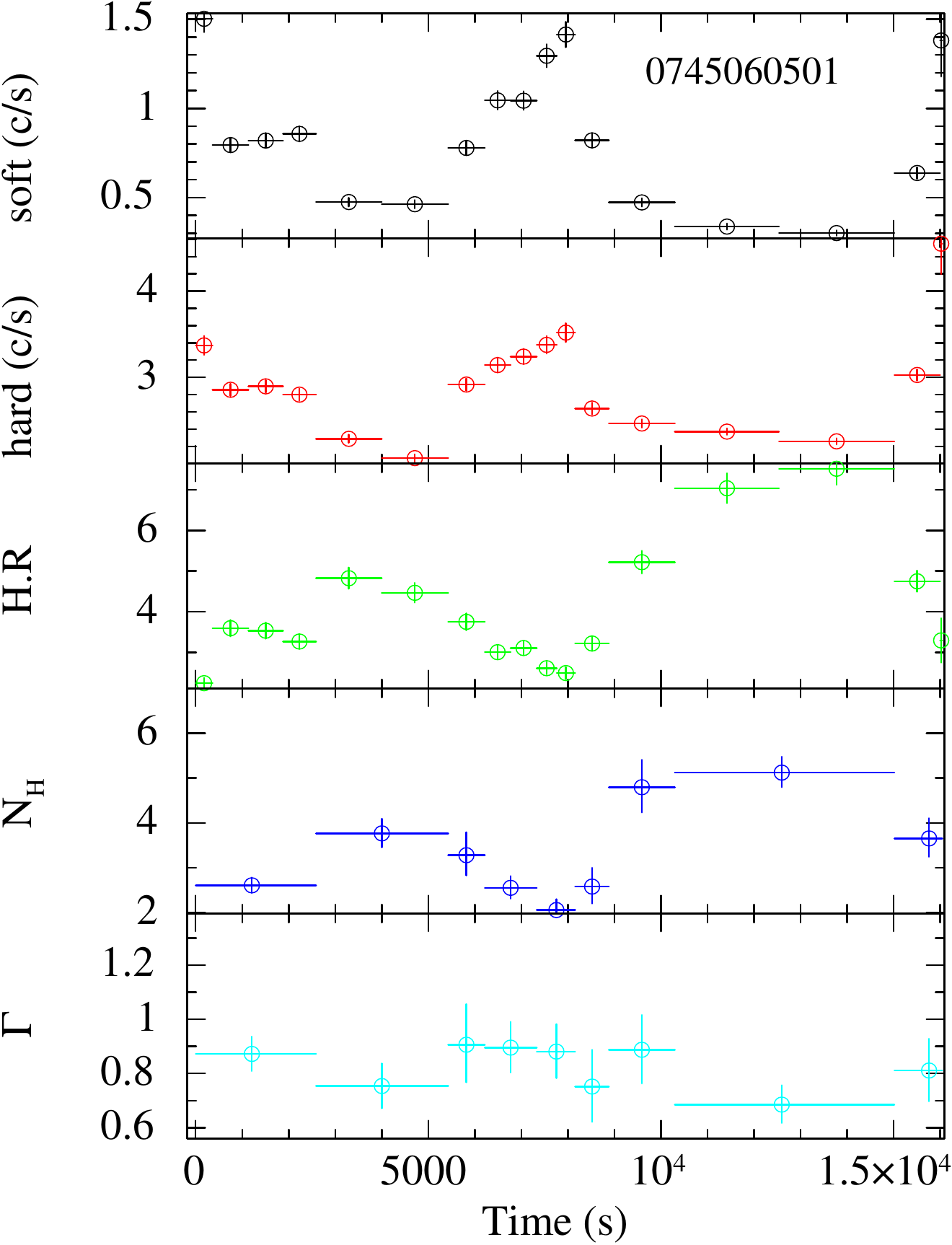}
 \includegraphics[width=6cm,height=7cm,angle=0]{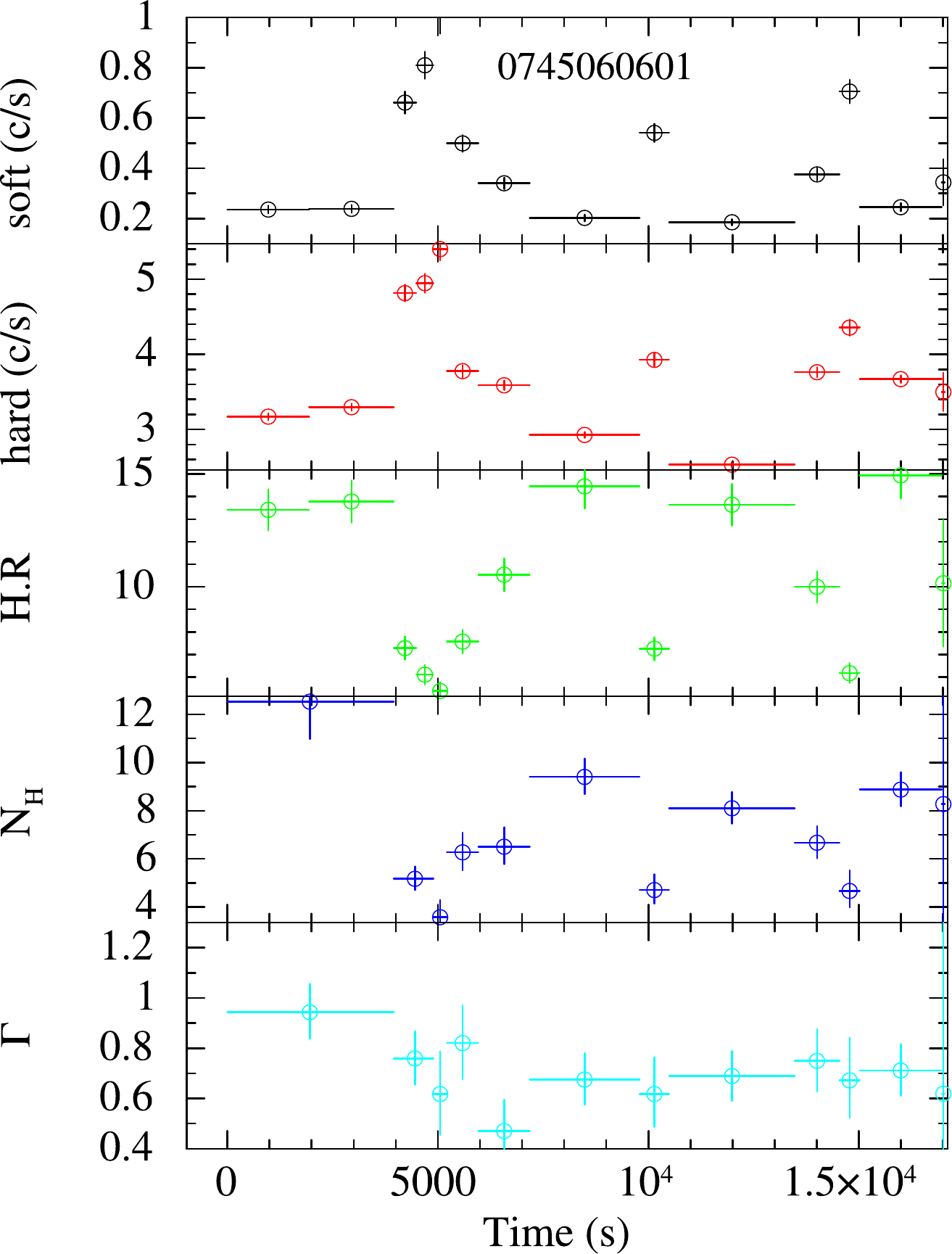}
 \includegraphics[width=6cm,height=7cm,angle=0]{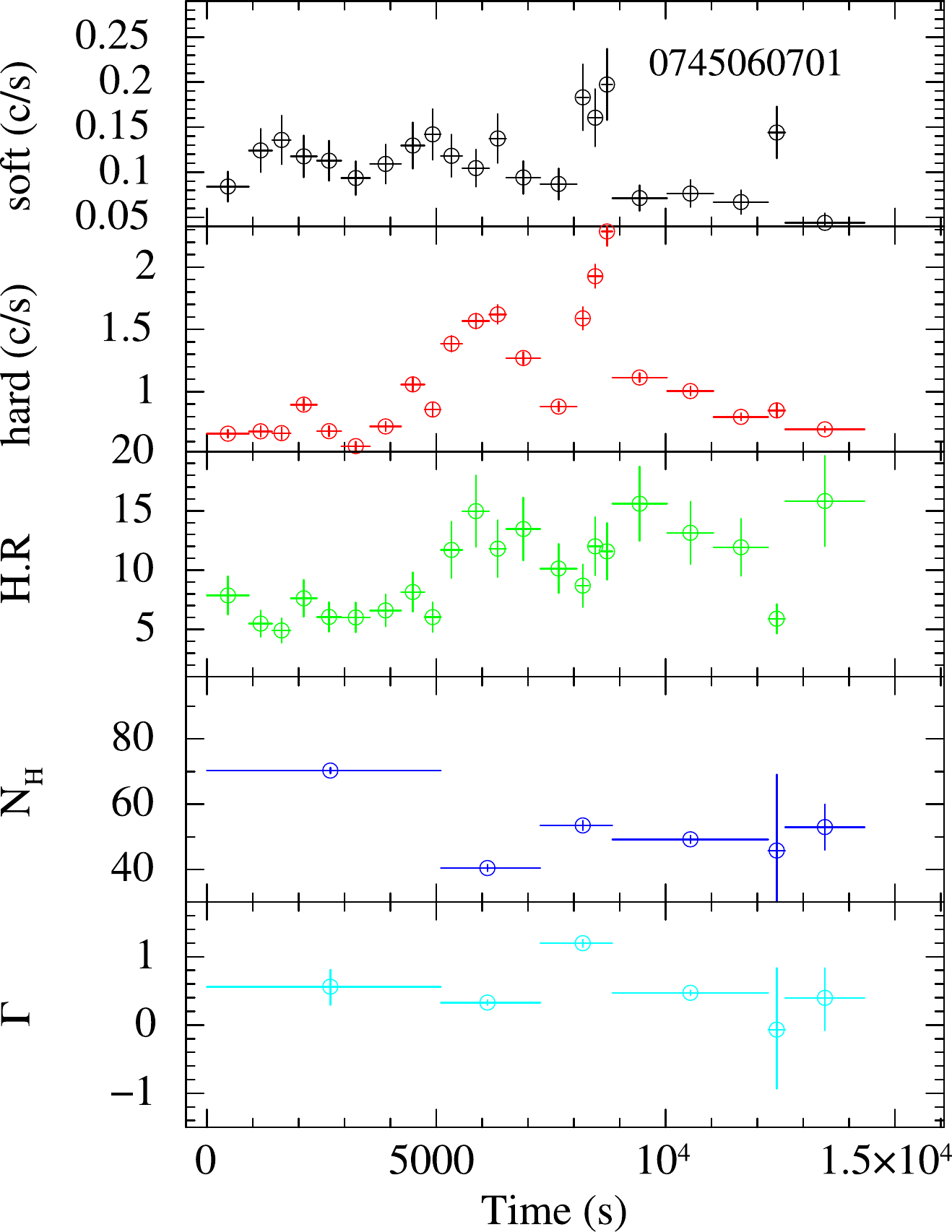}
 \includegraphics[width=6cm,height=7cm,angle=0]{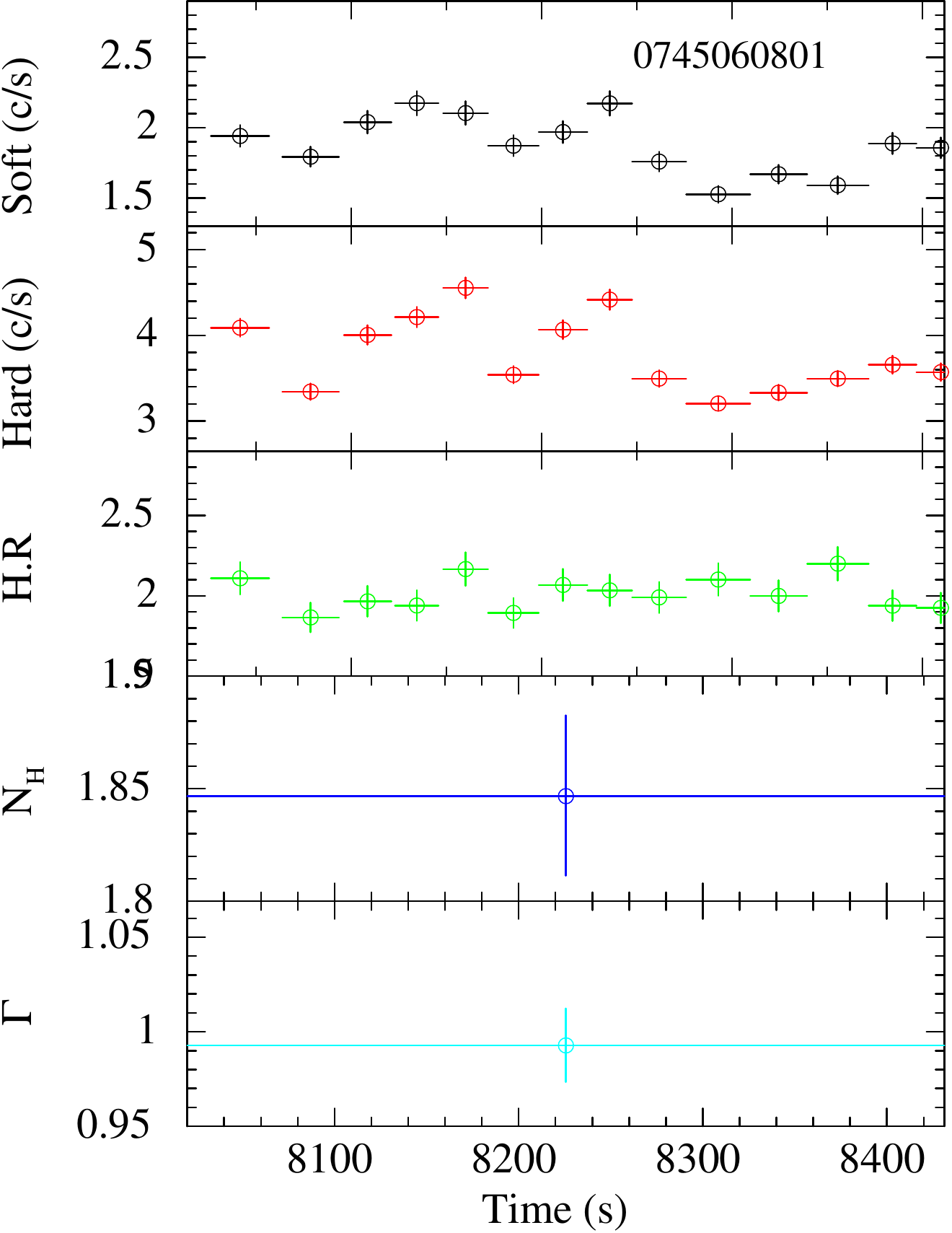}
 \caption{Results obtained from the HR-resolved spectral analysis performed on each \xmm\ observation. We show in each figure the adaptively-rebinned energy-resolved lightcurves, together with the correspondingly computed HR. We also show in blue and in cyan the values of the absorption column density and power-law photon index measured in the different time intervals that have been combined following the HR variations. The error bars on the X-axis in the $N_{\rm H}$ and the $\Gamma$ panels of each figure represents the extension in time of the combined time intervals with similar HR values used for the spectral extraction.}
\label{hr}
 \end{figure*}

We investigated the presence of spectral changes associated with the X-ray variability in each \xmm\ observation by extracting also the energy resolved lightcurves of the source with a time bin corresponding to the spin period measured above in each observation and computing the hardness ratio (HR). Based on our previous analysis on similar sources with the same technique \citep{bozzo2013, bozzo2017}, we selected the energy bands 0.5-3~keV and 3-10~keV, and computed the corresponding hardness ratio (HR) by using a further adaptive rebinning to achieve in each bin of the soft energy band lightcurve a minimum signal-to-noise ratio S/N$\gtrsim$10. The MOS and pn source spectra were then extracted by combining HR bins with similar values in each observation, following as closely as possible the observed HR variations. All these HR-resolved spectra were fit with an absorbed power-law model, as the statistics in all cases was not high enough to require the addition of the partial covering and the iron emission lines (see above in this section). The results of the HR-resolved spectral analysis are reported in Fig.~\ref{hr}. 

\section{Discussion}
\label{sec: results}

From the analysis of the average spectra extracted from the five \xmm\ observations carried out within the same NS revolution and described in Sect.~\ref{sec: data}, we found that the spectral energy distribution of the X-ray emission from \src\ in the \xmm\ energy band is similar to that of other classical SgXBs, being well described by a relatively hard power-law ($\Gamma\lesssim$1) and a partial covering component. 
In most of the spectra we also detected the iron K$\alpha$ and K$\beta$ line, as it is usually the case for a wind-fed classical SgXB where the X-rays from the NS give rise to fluorescence emission on the surrounding stellar wind material. The EW of the lines significantly increases during the eclipse, due to the depression of the X-ray continuum and the enhanced role of the fluorescence emission which is not (or only partly) occulted together with the NS. We checked during the fits to the different averaged spectra that the ratio between the normalization of the K$\alpha$ and K$\beta$ lines are compatible with the expected value of 0.15-0.16 \citep[see, e.g.][]{molendi03}. The results obtained from the isolated and older observation OBSID~0206380601 turned out to be very close to those determined during the last 46~ks of the OBSID~0745060801, as the two observations were carried out at a similar orbital phase (after the complete egress from the eclipse). 

Comparing the results of the fits to all average spectra, we conclude that the largest increases in the local absorption column density are measured close to the eclipse ingress and egress. When we look at Table~\ref{spec}, we notice that for those spectra where the statistics required a partial covering model and for which we tentatively distinguished in the spectral fit between a Galactic and local absorption component (OBSID~0745060401 - egress-, 0745060501, 0745060601, and 0745060701), the former remains virtually constant at $N_{\rm H1}$$\sim$2-3$\times$10$^{22}$~cm$^{-2}$, while the local absorption component seems to be constant around a value of $N_{\rm H2}$$\sim$6-7$\times$10$^{22}$~cm$^{-2}$ when we are sufficiently far from the eclipse ingress and egress. Among these higher statistics observations, a significant increase of $N_{\rm H2}$ is measured only during the OBSID~0745060401 (egress) and 0745060701, but data during these two observations were recorded at the earlier stages of the eclipse egress and when the source was almost completely eclipsed, respectively. Therefore, we can expect that the value of $N_{\rm H2}$ is significantly affected by the presence of the supergiant companion coming closer along the line of sight to the observer. For the OBSID~0206380601, where the statistics was too low to include a partial absorption column density, the measured value of the combined local+Galactic absorption matches relatively well the sum of $N_{\rm H1}$ and $N_{\rm H2}$ measured during the OBSID~0745060501 and 0745060601, supporting our suggestion that there could be no major increases in the local absorption column density in \src\ as a function of the orbital period. The only exception to this scenario is the OBSID~0745060801, which shows a low combined local+Galactic absorption. Given the fact that the lightcurve of this observation is particularly flat with no flares, we cannot exclude that during this particular observation (or orbital phase) the NS is accreting through a rather smooth portion of the wind with less clumps and more ionized material (this is likely to be only the material closer to the NS and thus not largely contributing to the production of the iron K$\alpha$ line the energy of which is still compatible with neutral iron distributed on a much larger region around the binary). Although the  coverage in orbital phase of the \xmm\ observations is relatively limited, this result is in agreement with what has been reported through the more extended (but less sensitive) orbital monitoring performed with \swift\,/XRT \citep{aftab2016}. Also in that case, the authors showed that the absorption column density in the direction to the source {measured from the spectra averaged in different orbital phase bins over the many different revolutions covered by the \swift\ satellite} is relatively constant away from the eclipse and shows minor fluctuations around an average value of $\sim$10$^{23}$~cm$^{-2}$. Looking at their Fig.~6, it seems also that they confirm a somewhat lower than average absorption during phases 0.17-0.25. This is at odds with respect at least to the case of another classical SgXBs for which an orbital monitoring has been carried out with \xmm,\ i.e. IGR\,J17252-3616 \citep{Manousakis2011}. In the case of this source, the authors found evidences for the presence of a massive absorbing structure placed between the NS and the observer at a specific orbital phase. This structure was associated with an accretion wake preceding the NS along its revolution around the companion. The formation of similar structures is usually ascribed to the presence of a slow stellar wind which can be inherently slow or slowed down by the photo-ionization produced by the NS X-ray emission \citep{Manousakis2012}. As extensively discussed in \citet[][see also references therein]{nunez17}, we are still missing global calculations that can take all complex parameters of the interaction between the massive star wind and the X-rays/gravitational pull from the compact object into account. Deriving absolute 
properties of the stellar wind (e.g. the terminal velocity) from X-ray observations of SgXBs and HMXBs in general is thus a complex exercise, results from which have to be taken with cautions \citep[see also][]{bozzo2011}.  However, as the X-ray flux measured by \xmm\ along the orbits of \src\ and IGR\,J17252-3616 are not too dissimilar (taking into account also the poorly known distance to both sources), it is possible that the stellar wind velocity in \src\ is somewhat higher than in IGR\,J17252-3616 (at least close to the compact object), and no accretion wake is formed. This suggestion remains so far speculative as we have no further means to draw a firm conclusion. It could however be used as a driver for interesting investigations of \src\ in different energy domains able to get more direct measurements of the stellar wind velocity at different orbital phases and possibly at different radial distances from the NS. We cannot exclude, as an alternative possibility, that some peculiar geometry of the accretion wake combined with the inclination of the line of sight to the observer could hamper the possibility of detecting such structure. So far, little is known from both a theoretical and observational perspective about the possible geometries of the accretion wakes in SgXBs and simulations still did not include different geometrical effects related, e.g., to the observer line of sight inclination and shape of the accretion wake to predict different observational outcomes \citep[see, e.g.,][and references therein]{Manousakis2012}.

Taking advantage of the large effective area and good energy resolution of the EPIC cameras, we could also study in details the spectral variability on much shorter time scales than the different orbital phases, i.e. within few hundreds to thousands seconds. The energy-resolved source lightcurves and the corresponding HR revealed that in each observation the source displayed a remarkable X-ray variability, as expected from a wind-fed classical SgXBs. Note that, being the lightcurves binned at the best determined source spin period in each observation, the observed variability in Fig.~\ref{lcurve} is not due to the effect of the NS pulsations. Based on our previous work on all \xmm\ observations of the SFXT sources, we performed an HR-resolved spectral analysis also for the \xmm\ observations of \src\ to probe the nature of the HR and the possibly associated spectra variability. \src\ is been used here as a test-bench to compare the results of a similar analysis performed on the two known sub-classes of SgXBs (classical systems and SFXTs) to help investigating possible systematic differences in the physical properties of the stellar wind from which the NS is accreting. From the plots in Fig.~\ref{hr}, we note a few interesting outcomes.  

The first is that there seems to be a clear decrease in the local absorption column density during the peaks of the observed  flares. This is particularly well noticeable for the flares recorded in the OBSID~0745060401 (during the egress from the eclipse), 0745060501, and 0745060601. The emission becomes softer at the peaks of the flares, and the results of the HR-resolved spectral analysis show that this is due to a decrease of the $N_{\rm H}$ rather than a change in the power-law photon index. The same behavior was observed in the case of most of the SFXT sources \citep{bozzo2017} and it was ascribed to the fact that the higher flux during the peak of a flare can photoionize the surrounding material from which the NS is accreting and produce a rapid decrease of the local absorption column density. The increase of the $N_{\rm H}$ following the peak of the flare can be explained as being due to the recombination in the wind material during the decrease of the X-ray flux along the decay of the flare. As the photoionization (and thus the decrease in the absorption column density) increases with the X-ray flux, it is interesting to note that this effect is much less pronounced during the flares recorded in the OBSID~0745060701 which reached a substantially lower count-rate at the peak compared to the flares in OBSID~0745060501 and 0745060601. 

In order to provide support to the fact that flares are triggered by the presence of clumps, in the case of the SFXTs it was highlighted that often the rise of flares is accompanied by an increase in the local absorption column density before the onset of a flare  \citep{bozzo2017}. This can be interpreted as being due to the clump approaching the NS and contributing to an enhancement of the local absorbing material. A similar effect seems to be present in the case of \src,\ as we can see from the HR-resolved spectral analysis during the OBSID~0745060501 (time intervals 3-5~ks and 11-15~ks after the beginning of the observation), 0745060601 (time intervals 0-4~ks, 7-10~ks, and 11-14~ks after the beginning of the observation), and 0745060701 (time intervals 0-5~ks  and 7-9~ks after the beginning of the observation). 

The OBSID~0206380601 and 0745060801 are characterized by a much reduced (if any at all) HR variations. While for the OBSID~0206380601 it is difficult to achieve a firm conclusion due to the low statistics and the fragmented lightcurve caused by the presence of several high-flaring background time intervals (see Sect.~\ref{sec: data}), the situation for the OBSID~0745060801 is clearer. As discussed previously in this section, the low absorption column density measured during this observation and the less pronounced variability compared to the other monitored orbital phases could be caused by an higher ionization of a somewhat lower density wind material close to the compact object and/or to the lack of the presence of clumps. The similarities (to be confirmed) in the X-ray emission properties of the source in the OBSID~0206380601 and 0745060801 suggests that during the orbital phase $\sim$0.17-0.25 the NS might be regularly accreting from a less structured wind material. At present, there are no suitable data to verify further this hypothesis, as the only other available dataset is the one published by \citet{aftab2016} but the \swift\/XRT lightcurves are too fragmented and endowed with a too low statistics to perform any meaningful HR-resolved analysis. 
Future observations at these orbital phases with \xmm\ might be able to confirm our proposed interpretation or help finding different explanations.

\section{Conclusion}
\label{sec: disc}

In this paper we exploited the available \xmm\ observations of \src\ in order to carry out for the first time the HR-resolved spectral analysis of the X-ray emission from a classical SgXB and compare the findings with those obtained from a similar analysis performed systematically on all \xmm\ observations of the SFXTs \citep{bozzo2017}. Taking advantage of the good energy resolution and large effective area of the EPIC cameras on-board \xmm,\ the HR-resolved spectral analysis could be carried out in all cases on timescales much shorter than the duration of the flares and could thus be used to search for fast spectral variations that can reveal if similar or different mechanisms are triggering these events in classical SgXBs as \src\ and in the SFXTs. 

From the results summarized in Fig.~\ref{hr} and the comments presented in Sect.~\ref{sec: results}, we concluded that also in the case of the classical SgXB \src,\ the spectral variations measured during flares are compatible with the idea that these events are triggered by the presence of clumps. We generally observed that the local absorption column density increases during the rises to the flares, suggesting that a dense structure is approaching the NS and giving rise to an enhanced mass accretion rate. At the peak of the flares, a decrease of the absorption column density is usually measured, compatible with the idea that the enhanced X-ray flux is able to photoionize the clump material. Recombination is likely to be fast during the decay of the flare \citep[see, e.g., the discussion in][and references therein]{bozzo2011}, and thus the increase of the local absorption column density that is measured after the peak of the flares can be ascribed to an increasingly lower photoionization state of the wind material while part of this clump has been accreted and the remainings move away from the compact object. 


The present analysis of \src\ indicates that wind clumps are among the causes of the onset of flares in this classical SgXB, 
similarly to what was found for SFXTs. At present, we cannot therefore firmly establish if the different behavior observed from SFXTs compared to \src\ 
(and more generally to the whole class of classical SgXBs) is only due to a systematic difference between the clump properties in these two classes of objects. 
A more systematic analysis of the \xmm\ observations performed in the direction in all other observed classical SgXBs is underway and will be reported in a 
forthcoming publication.  

\section*{acknowledgements} 
The data for this work has been obtained through the High Energy Astrophysics Science Archive (HEASARC) Online Service provided by NASA/GSFC. 
We would also like to thank the anonymous referee for his/her invaluable comments and suggestions.

\appendix

\section{ \xmm\ energy-resolved pulse profiles of \src}
\label{sec:appendix}

We report in this section the energy-resolved pulse profiles of \src,\ as determined from all \xmm\ observations in Table~\ref{tab: log}. The pulse profiles have been obtained from each observation folding the extracted lightcurves in the 0.5-3~keV and 3-10~keV energy bands (see Fig.~\ref{lcurve}) on the best determined source pulse period in each observation. As it can be seen from the figures below, the pulse profiles are characterized by two main peaks, in agreement with what was found by \citet{aftab2016}, and we refer the reader also to this paper for a discussion on the X-ray pulse profiles from \src.\ We notice that the complex shapes of the profiles and the variations from one observation to the other is qualitatively compatible with what is usually observed from high mass X-ray binaries (HMXBs) hosting strongly magnetized NSs \citep[see, e.g.,][]{staubert}. In these sources, the complexity of the pulse profiles and their changes as a function of time and energy are associated to the topology of the magnetic field and the formation of extended accretion columns within which the kinetic and potential energy of the accreting matter is converted in X-ray radiation. Providing an interpretation of the specific pulse profile shapes observed from \src\ is beyond the scope of the present paper, but we notice that all previous attempts to invert the profiles of HMXBs to derive constraints on the magnetic field configuration of the NS proved so far very challenging and advanced modelling studies are on-going \citep[see, e.g.,][and references therein]{becker_wolf2007,hmxb1,hmxb2,becker2018}.  
\begin{figure*}
 \centering
 \includegraphics[width=8cm,angle=0]{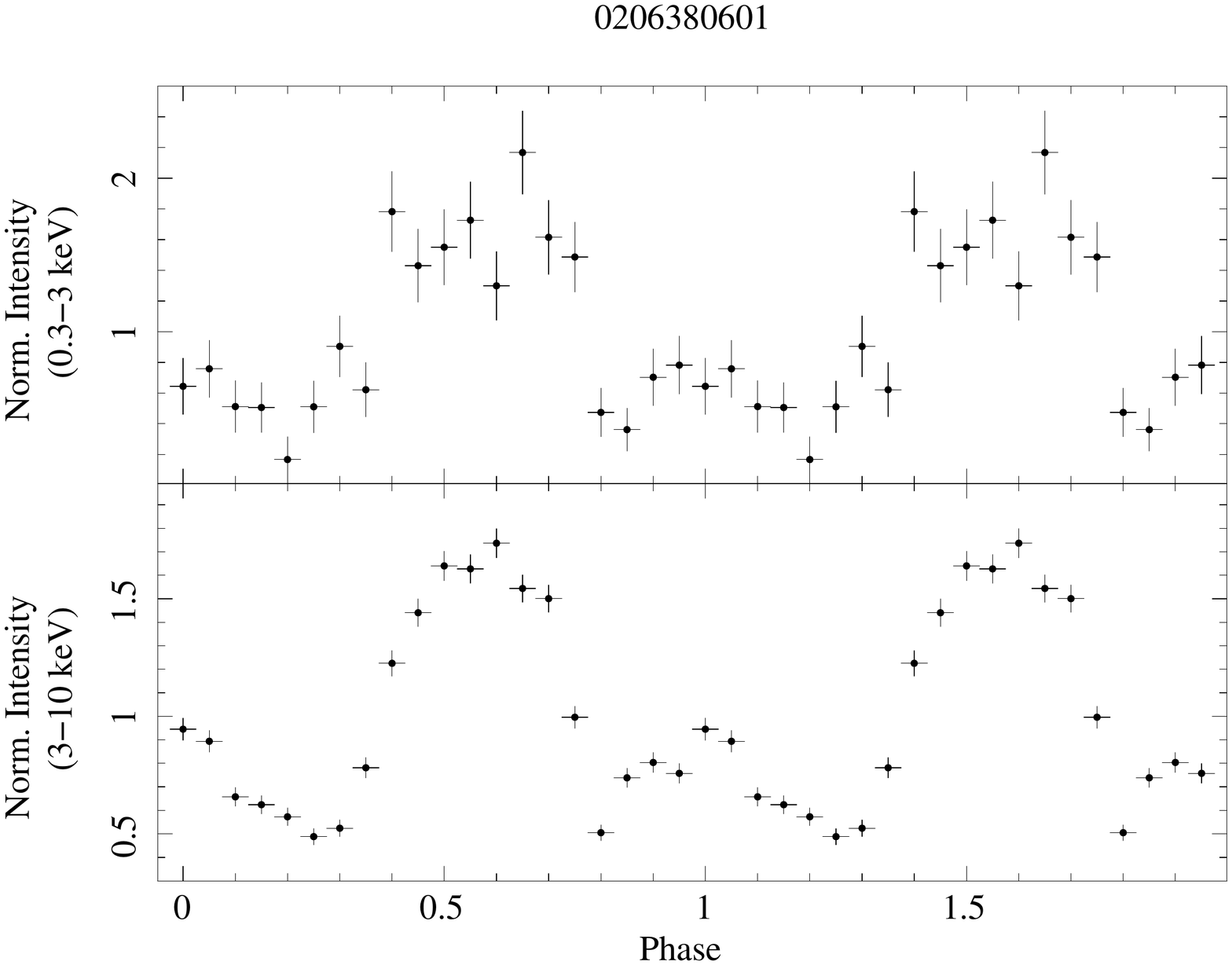}
 \includegraphics[width=8cm,angle=0]{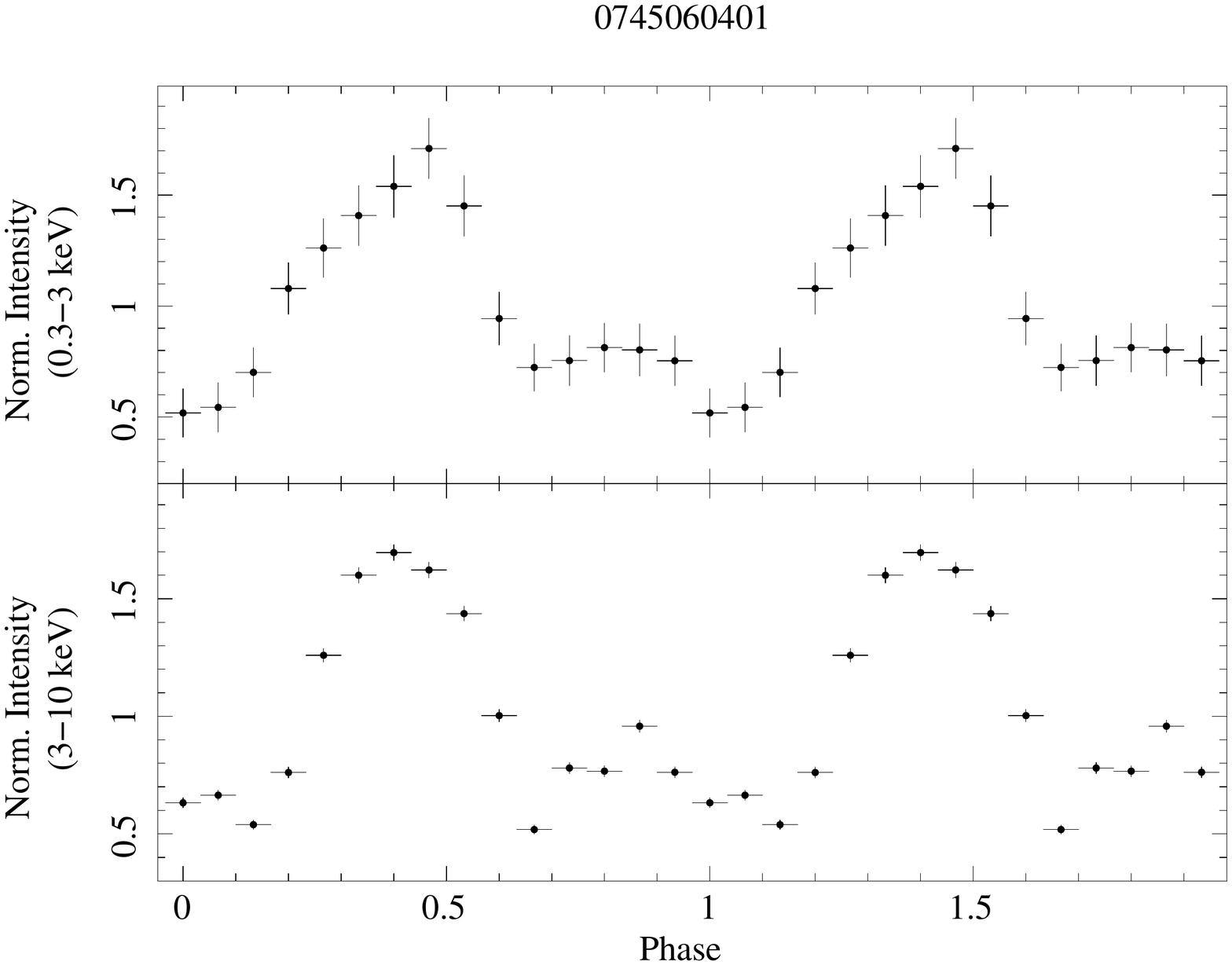} 
 \includegraphics[width=8cm,angle=0]{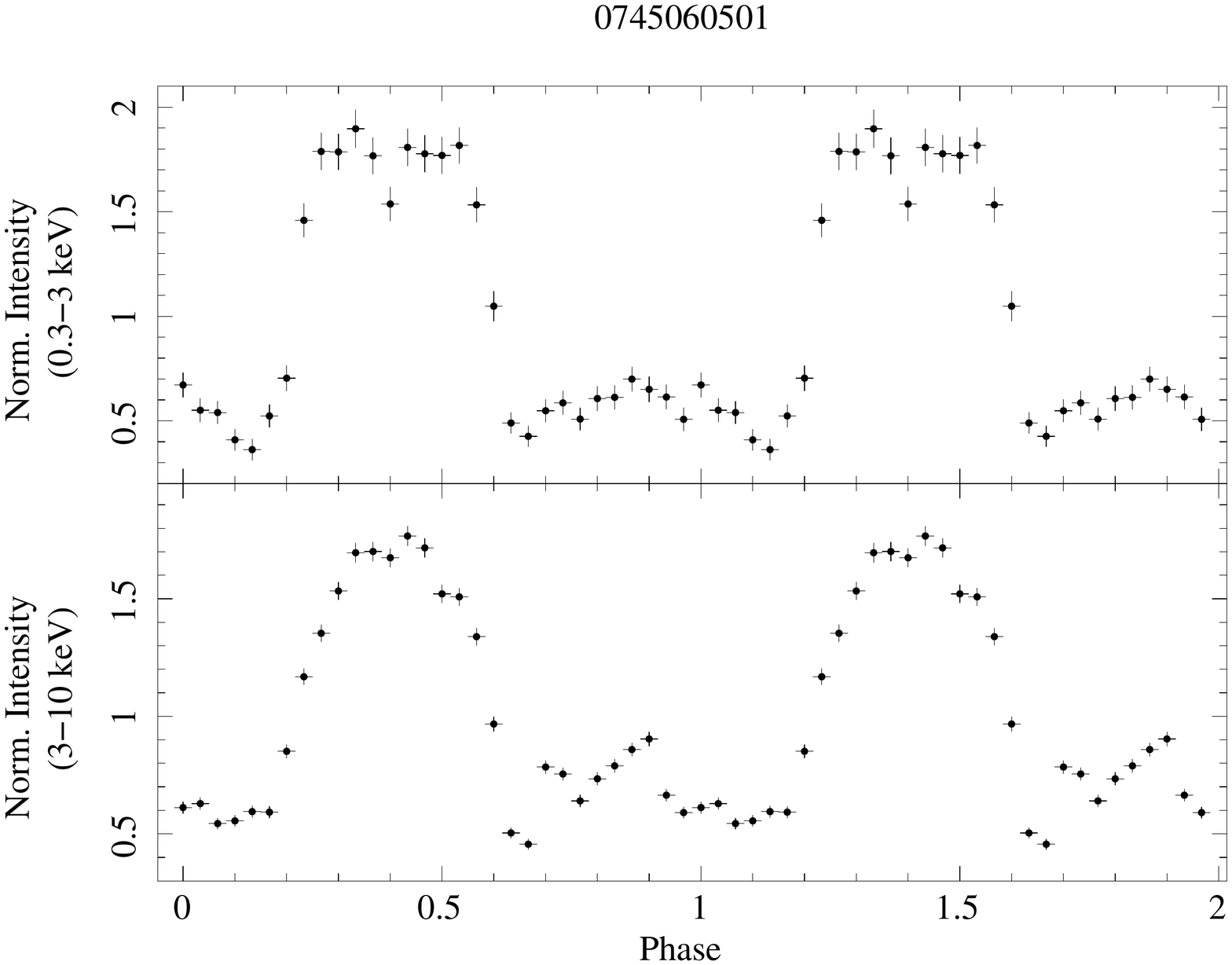}
 \includegraphics[width=8cm,angle=0]{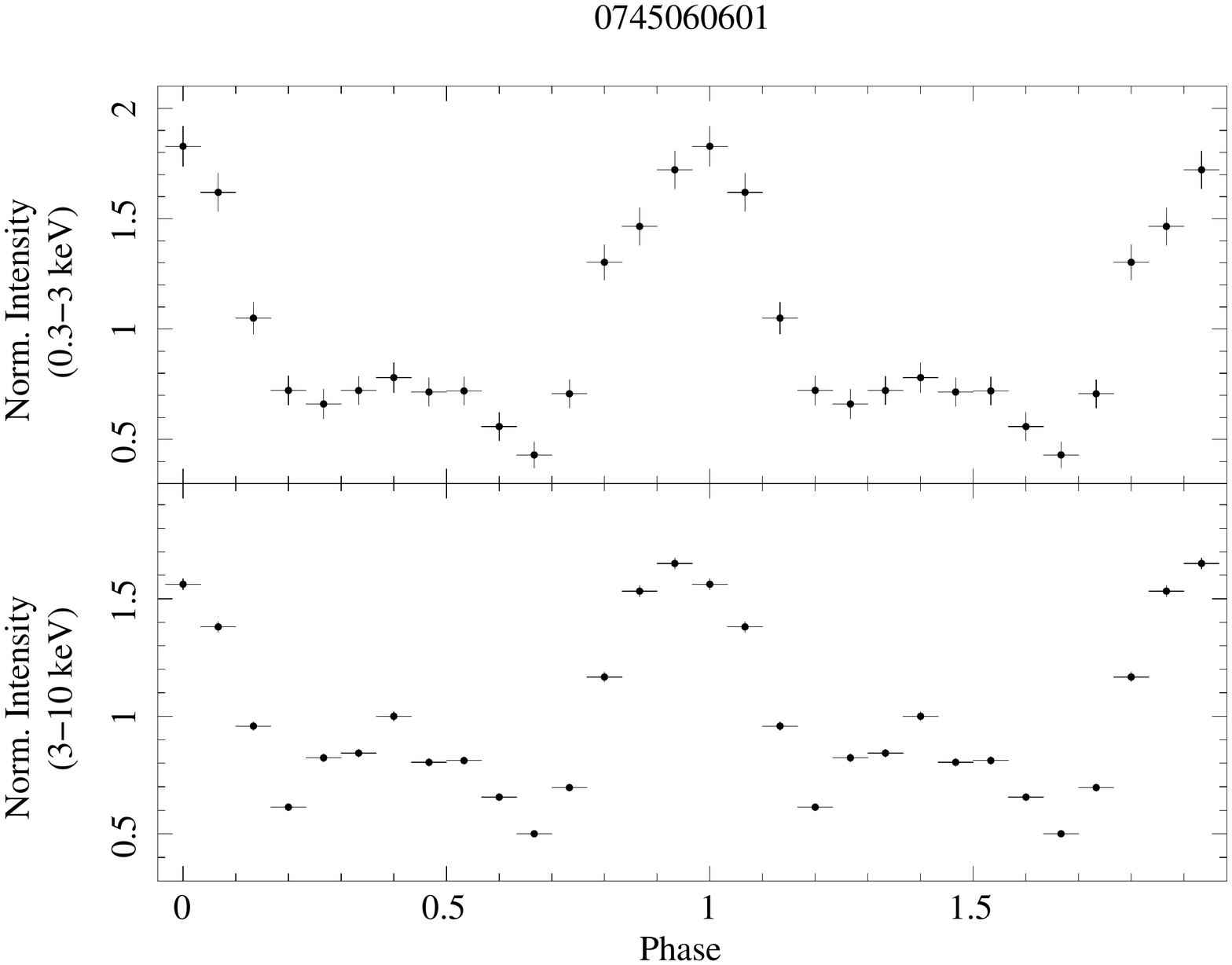}
 \includegraphics[width=8cm,angle=0]{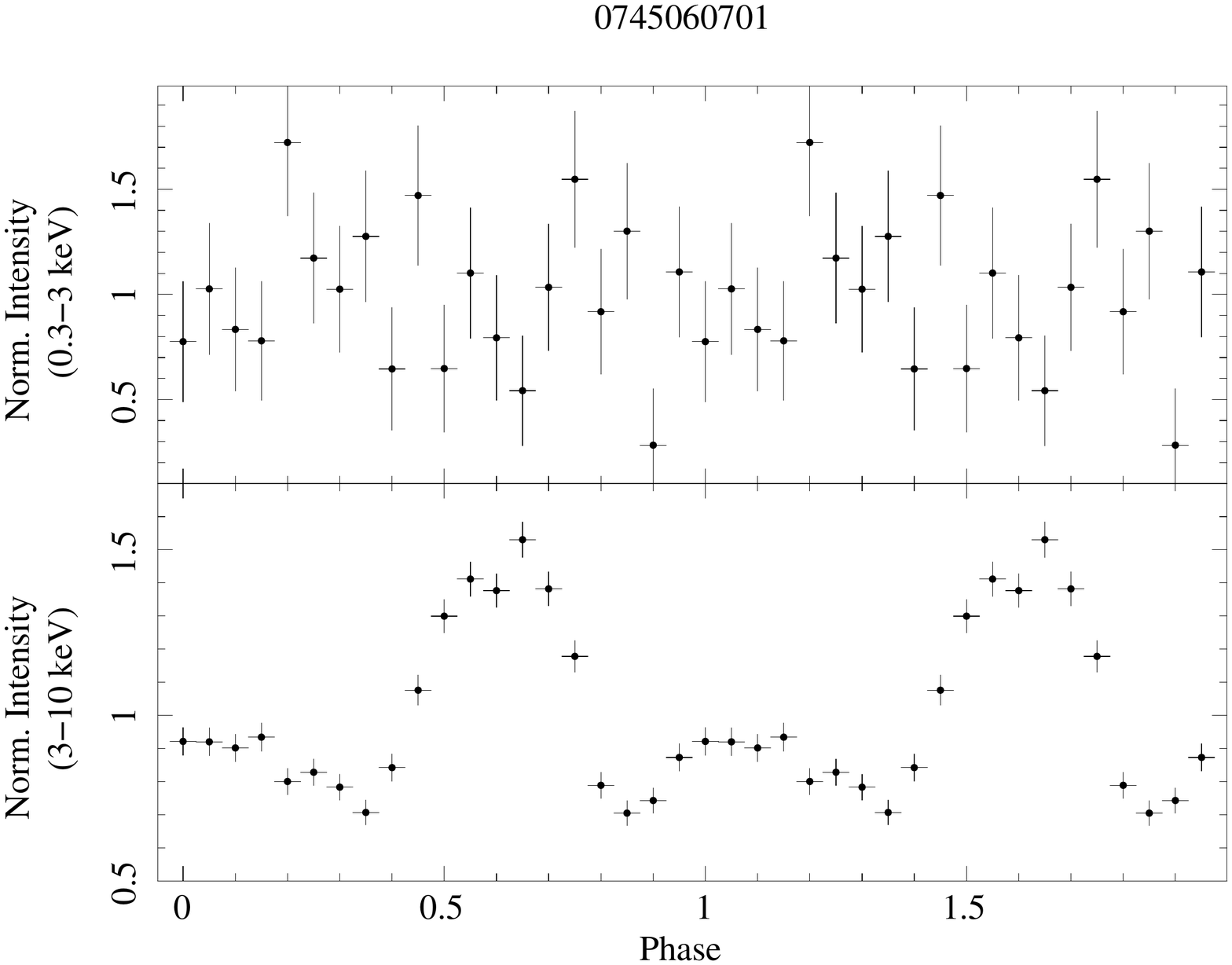}
 \includegraphics[width=8cm,angle=0]{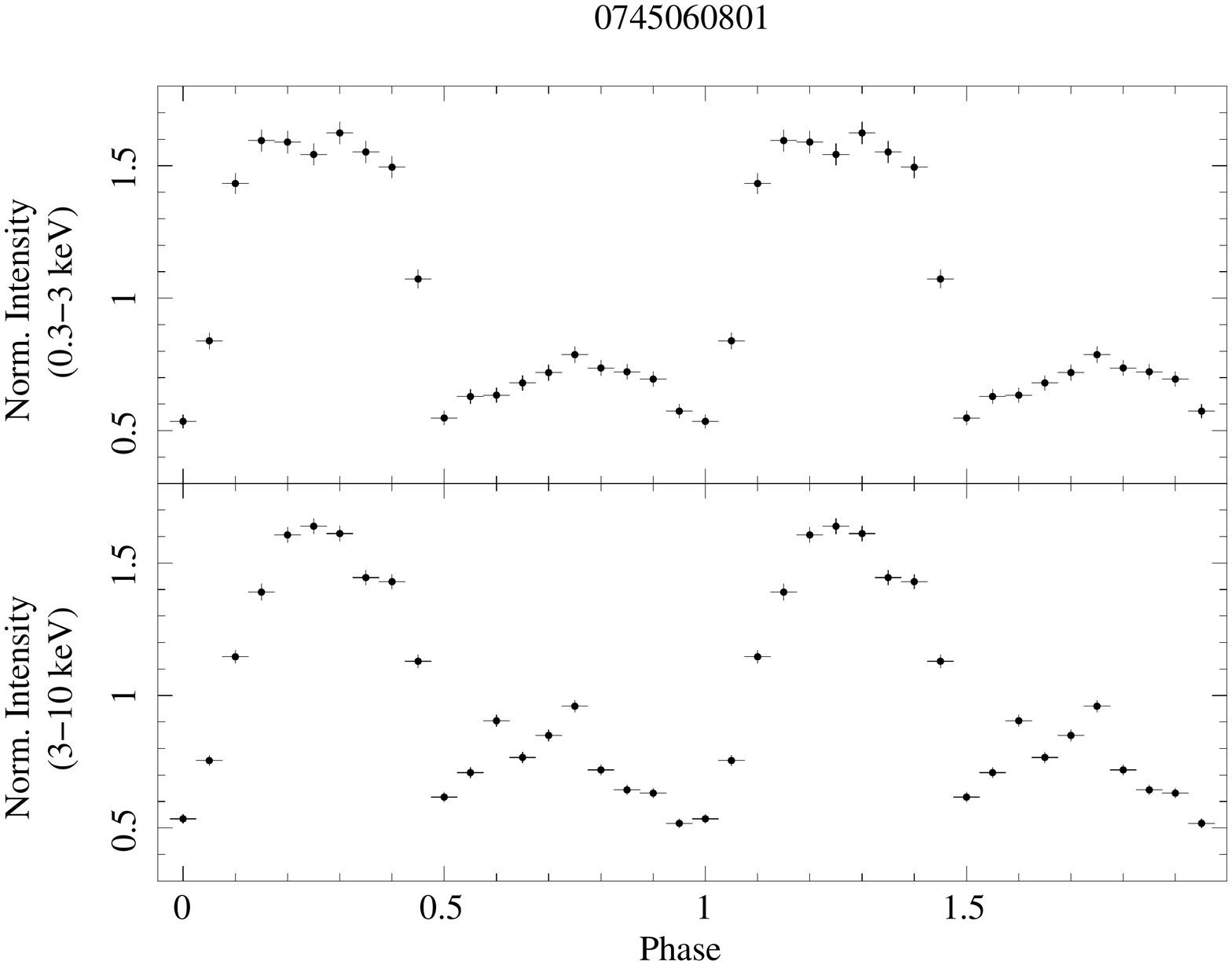}
 \caption{\xmm\ energy-resolved pulse profiles of \src.\ The two peaks, as reported previously by \citet{aftab2016}, are visible in all observations. Note that the observation ID~0745060701 caught the source during the ingress to the eclipse, and the large absorption column density hampered the detection of pulsations in the 0.3-3~keV energy band.}
\label{profiles}
 \end{figure*}

\bibliographystyle{pwterse-ay}
\bibliography{reference}

\begin{thebibliography}{51}
\providecommand\natexlab[1]{#1}
\providecommand\JournalTitle[1]{#1}

\bibitem[{{Aftab} et~al.(2016){Aftab}, {Islam}, \& {Paul}}]{aftab2016}
{Aftab}~N, et~al.
\href{http://dx.doi.org/10.1093/mnras/stw2037}{2016 \JournalTitle{\mnras} 463
  2032}.

\bibitem[{{Anders} \& {Grevesse}(1989)}]{anders1989}
{Anders}~E, \& {Grevesse}~N.
\href{http://dx.doi.org/10.1016/0016-7037(89)90286-X}{1989 \JournalTitle{\gca}
  53 197}.

\bibitem[{{Becker} \& {Wolff}(2007)}]{becker_wolf2007}
{Becker}~PA, \& {Wolff}~MT.
\href{http://dx.doi.org/10.1086/509108}{2007 \JournalTitle{\apj} 654 435}.

\bibitem[{{Becker} \& {Wolff}(2018)}]{becker2018}
{Becker}~PA, \& {Wolff}~MT.
2018 in American Astronomical Society Meeting Abstracts, Vol. 231, American
  Astronomical Society Meeting Abstracts \#231 333.04.

\bibitem[{{Bozzo} et~al.(2017){Bozzo}, {Bernardini}, {Ferrigno}, {Falanga},
  {Romano}, \& {Oskinova}}]{bozzo2017}
{Bozzo}~E, et~al.
\href{http://dx.doi.org/10.1051/0004-6361/201730398}{2017 \JournalTitle{\aap}
  608 A128}.

\bibitem[{{Bozzo} et~al.(2008){Bozzo}, {Falanga}, \& {Stella}}]{bozzo2008}
{Bozzo}~E, et~al.
\href{http://dx.doi.org/10.1086/589990}{2008 \JournalTitle{\apj} 683 1031}.

\bibitem[{{Bozzo} et~al.(2015){Bozzo}, {Romano}, {Ducci}, {Bernardini}, \&
  {Falanga}}]{bozzo2015}
{Bozzo}~E, et~al.
\href{http://dx.doi.org/10.1016/j.asr.2014.11.012}{2015 \JournalTitle{Advances
  in Space Research} 55 1255}.

\bibitem[{{Bozzo} et~al.(2013{\natexlab{a}}){Bozzo}, {Romano}, {Ferrigno},
  {Campana}, {Falanga}, {Israel}, {Walter}, \& {Stella}}]{bozzo13}
{Bozzo}~E, et~al.
\href{http://dx.doi.org/10.1051/0004-6361/201321248}{2013{\natexlab{a}}
  \JournalTitle{\aap} 556 A30}.

\bibitem[{{Bozzo} et~al.(2013{\natexlab{b}}){Bozzo}, {Romano}, {Ferrigno},
  {Campana}, {Falanga}, {Israel}, {Walter}, \& {Stella}}]{bozzo2013}
{Bozzo}~E, et~al.
\href{http://dx.doi.org/10.1051/0004-6361/201321248}{2013{\natexlab{b}}
  \JournalTitle{\aap} 556 A30}.

\bibitem[{{Bozzo} et~al.(2011){Bozzo}, {Giunta}, {Cusumano}, {Ferrigno},
  {Walter}, {Campana}, {Falanga}, {Israel}, \& {Stella}}]{bozzo2011}
{Bozzo}~E, et~al.
\href{http://dx.doi.org/10.1051/0004-6361/201116726}{2011 \JournalTitle{\aap}
  531 A130}.

\bibitem[{{Bozzo} et~al.(2016){Bozzo}, {Bhalerao}, {Pradhan}, {Tomsick},
  {Romano}, {Ferrigno}, {Chaty}, {Oskinova}, {Manousakis}, \&
  {Walter}}]{bozzo2016}
{Bozzo}~E, et~al.
\href{http://dx.doi.org/10.1051/0004-6361/201629311}{2016 \JournalTitle{\aap}
  596 A16}.

\bibitem[{{Chaty} et~al.(2008){Chaty}, {Rahoui}, {Foellmi}, {Tomsick},
  {Rodriguez}, \& {Walter}}]{chaty08}
{Chaty}~S, et~al.
\href{http://dx.doi.org/10.1051/0004-6361:20078768}{2008 \JournalTitle{\aap}
  484 783}.

\bibitem[{{Coley} et~al.(2015){Coley}, {Corbet}, \& {Krimm}}]{coley2015}
{Coley}~JB, et~al.
\href{http://dx.doi.org/10.1088/0004-637X/808/2/140}{2015 \JournalTitle{\apj}
  808 140}.

\bibitem[{{Drave} et~al.(2014){Drave}, {Bird}, {Sidoli}, {Sguera}, {Bazzano},
  {Hill}, \& {Goossens}}]{drave2014}
{Drave}~SP, et~al.
\href{http://dx.doi.org/10.1093/mnras/stu110}{2014 \JournalTitle{\mnras} 439
  2175}.

\bibitem[{{Ducci} et~al.(2013){Ducci}, {Romano}, {Esposito}, {Bozzo}, {Krimm},
  {Vercellone}, {Mangano}, \& {Kennea}}]{ducci2013}
{Ducci}~L, et~al.
\href{http://dx.doi.org/10.1051/0004-6361/201321635}{2013 \JournalTitle{\aap}
  556 A72}.

\bibitem[{{Ducci} et~al.(2009){Ducci}, {Sidoli}, {Mereghetti}, {Paizis}, \&
  {Romano}}]{ducci2009}
{Ducci}~L, et~al.
\href{http://dx.doi.org/10.1111/j.1365-2966.2009.15265.x}{2009
  \JournalTitle{\mnras} 398 2152}.

\bibitem[{{Falanga} et~al.(2015){Falanga}, {Bozzo}, {Lutovinov},
  {Bonnet-Bidaud}, {Fetisova}, \& {Puls}}]{falanga2015}
{Falanga}~M, et~al.
\href{http://dx.doi.org/10.1051/0004-6361/201425191}{2015 \JournalTitle{\aap}
  577 A130}.

\bibitem[{{Farinelli} et~al.(2016){Farinelli}, {Ferrigno}, {Bozzo}, \&
  {Becker}}]{hmxb2}
{Farinelli}~R, et~al.
\href{http://dx.doi.org/10.1051/0004-6361/201527257}{2016 \JournalTitle{\aap}
  591 A29}.

\bibitem[{{F{\"u}rst} et~al.(2010){F{\"u}rst}, {Kreykenbohm}, {Pottschmidt},
  {Wilms}, {Hanke}, {Rothschild}, {Kretschmar}, {Schulz}, {Huenemoerder},
  {Klochkov}, \& {Staubert}}]{fuerst2010}
{F{\"u}rst}~F, et~al.
\href{http://dx.doi.org/10.1051/0004-6361/200913981}{2010 \JournalTitle{\aap}
  519 A37}.

\bibitem[{{F{\"u}rst} et~al.(2014){F{\"u}rst}, {Pottschmidt}, {Wilms},
  {Tomsick}, {Bachetti}, {Boggs}, {Christensen}, {Craig}, {Grefenstette},
  {Hailey}, {Harrison}, {Madsen}, {Miller}, {Stern}, {Walton}, \&
  {Zhang}}]{velax1_furst2014}
{F{\"u}rst}~F, et~al.
\href{http://dx.doi.org/10.1088/0004-637X/780/2/133}{2014 \JournalTitle{\apj}
  780 133}.

\bibitem[{{Hill} et~al.(2005){Hill}, {Walter}, {Knigge}, {Bazzano},
  {B{\'e}langer}, {Bird}, {Dean}, {Galache}, {Malizia}, {Renaud}, {Stephen}, \&
  {Ubertini}}]{hill2005}
{Hill}~AB, et~al.
\href{http://dx.doi.org/10.1051/0004-6361:20052937}{2005 \JournalTitle{\aap}
  439 255}.

\bibitem[{{Iaria} et~al.(2004){Iaria}, {Augello}, {Robba}, {di Salvo},
  {Burderi}, \& {Lavagetto}}]{iaria2004R}
{Iaria}~R, et~al.
2004 in Revista Mexicana de Astronomia y Astrofisica Conference Series,
  Vol.~20, Revista Mexicana de Astronomia y Astrofisica Conference Series, ed.
  {Tovmassian}~G, \& {Sion}~E 212.

\bibitem[{{in't Zand}(2005)}]{intzand2005_sfxts}
{in't Zand}~JJM.
\href{http://dx.doi.org/10.1051/0004-6361:200500162}{2005 \JournalTitle{\aap}
  441 L1}.

\bibitem[{{Jain} et~al.(2009){Jain}, {Paul}, \& {Dutta}}]{jain2009}
{Jain}~C, et~al.
\href{http://dx.doi.org/10.1088/1674-4527/9/12/002}{2009 \JournalTitle{Research
  in Astronomy and Astrophysics} 9 1303}.

\bibitem[{{Kreykenbohm} et~al.(2008){Kreykenbohm}, {Wilms}, {Kretschmar},
  {Torrej{\'o}n}, {Pottschmidt}, {Hanke}, {Santangelo}, {Ferrigno}, \&
  {Staubert}}]{Kreykenbohm2008}
{Kreykenbohm}~I, et~al.
\href{http://dx.doi.org/10.1051/0004-6361:200809956}{2008 \JournalTitle{\aap}
  492 511}.

\bibitem[{{Krti{\v c}ka} \& {Kub{\'a}t}(2016)}]{kri16}
{Krti{\v c}ka}~J, \& {Kub{\'a}t}~J.
\href{http://dx.doi.org/10.1016/j.asr.2016.01.002}{2016 \JournalTitle{Advances
  in Space Research} 58 710}.

\bibitem[{{Krti{\v c}ka} et~al.(2015){Krti{\v c}ka}, {Kub{\'a}t}, \& {Krti{\v
  c}kov{\'a}}}]{kri15}
{Krti{\v c}ka}~J, et~al.
\href{http://dx.doi.org/10.1051/0004-6361/201525637}{2015 \JournalTitle{\aap}
  579 A111}.

\bibitem[{{Leahy}(1987)}]{fold}
{Leahy}~DA.
1987 \JournalTitle{\aap} 180 275.

\bibitem[{{Manousakis} \& {Walter}(2011)}]{Manousakis2011}
{Manousakis}~A, \& {Walter}~R.
\href{http://dx.doi.org/10.1051/0004-6361/201015707}{2011 \JournalTitle{\aap}
  526 A62}.

\bibitem[{{Manousakis} et~al.(2012){Manousakis}, {Walter}, \&
  {Blondin}}]{Manousakis2012}
{Manousakis}~A, et~al.
\href{http://dx.doi.org/10.1051/0004-6361/201219717}{2012 \JournalTitle{\aap}
  547 A20}.

\bibitem[{{Mart{\'{\i}}nez-N{\'u}{\~n}ez}
  et~al.(2017){Mart{\'{\i}}nez-N{\'u}{\~n}ez}, {Kretschmar}, {Bozzo},
  {Oskinova}, {Puls}, {Sidoli}, {Sundqvist}, {Blay}, {Falanga}, {F{\"u}rst},
  {G{\'{\i}}menez-Garc{\'{\i}}a}, {Kreykenbohm}, {K{\"u}hnel}, {Sander},
  {Torrej{\'o}n}, \& {Wilms}}]{nunez17}
{Mart{\'{\i}}nez-N{\'u}{\~n}ez}~S, et~al.
\href{http://dx.doi.org/10.1007/s11214-017-0340-1}{2017 \JournalTitle{\ssr}},
  \href{http://arxiv.org/abs/1701.08618}{{\sffamily arXiv:1701.08618
  [astro-ph.HE]}}.

\bibitem[{{Molendi} et~al.(2003){Molendi}, {Bianchi}, \& {Matt}}]{molendi03}
{Molendi}~S, et~al.
\href{http://dx.doi.org/10.1046/j.1365-8711.2003.06783.x}{2003
  \JournalTitle{\mnras} 343 L1}.

\bibitem[{{Negueruela} et~al.(2008){Negueruela}, {Torrej{\'o}n}, {Reig},
  {Rib{\'o}}, \& {Smith}}]{negueruela2008}
{Negueruela}~I, et~al.
\href{http://dx.doi.org/10.1063/1.2945052}{2008 in American Institute of
  Physics Conference Series, Vol. 1010, A Population Explosion: The Nature \&
  Evolution of X-ray Binaries in Diverse Environments, ed. {Bandyopadhyay}~RM,
  et~al.} 252.

\bibitem[{{Pradhan} et~al.(2014){Pradhan}, {Maitra}, {Paul}, {Islam}, \&
  {Paul}}]{pradhan2014}
{Pradhan}~P, et~al.
\href{http://dx.doi.org/10.1093/mnras/stu1034}{2014 \JournalTitle{\mnras} 442
  2691}.

\bibitem[{{Puls} et~al.(2008){Puls}, {Vink}, \& {Najarro}}]{puls08}
{Puls}~J, et~al.
\href{http://dx.doi.org/10.1007/s00159-008-0015-8}{2008 \JournalTitle{\aapr} 16
  209}.

\bibitem[{{Rampy} et~al.(2009){Rampy}, {Smith}, \& {Negueruela}}]{rampy2009}
{Rampy}~RA, et~al.
\href{http://dx.doi.org/10.1088/0004-637X/707/1/243}{2009 \JournalTitle{\apj}
  707 243}.

\bibitem[{{Revnivtsev} \& {Mereghetti}(2015)}]{rev15}
{Revnivtsev}~M, \& {Mereghetti}~S.
\href{http://dx.doi.org/10.1007/s11214-014-0123-x}{2015 \JournalTitle{\ssr} 191
  293}.

\bibitem[{{Revnivtsev} et~al.(2004){Revnivtsev}, {Sunyaev}, {Varshalovich},
  {Zheleznyakov}, {Cherepashchuk}, {Lutovinov}, {Churazov}, {Grebenev}, \&
  {Gilfanov}}]{rev2004}
{Revnivtsev}~MG, et~al.
\href{http://dx.doi.org/10.1134/1.1764884}{2004 \JournalTitle{Astronomy
  Letters} 30 382}.

\bibitem[{{Romano} et~al.(2015){Romano}, {Bozzo}, {Mangano}, {Esposito},
  {Israel}, {Tiengo}, {Campana}, {Ducci}, {Ferrigno}, \&
  {Kennea}}]{romano2015_17544}
{Romano}~P, et~al.
\href{http://dx.doi.org/10.1051/0004-6361/201525749}{2015 \JournalTitle{\aap}
  576 L4}.

\bibitem[{{Sako} et~al.(2003){Sako}, {Kahn}, {Paerels}, {Liedahl}, {Watanabe},
  {Nagase}, \& {Takahashi}}]{sako2003}
{Sako}~M, et~al.
2003 \JournalTitle{ArXiv Astrophysics e-prints},
  \href{http://arxiv.org/abs/astro-ph/0309503}{{\sffamily astro-ph/0309503}}.

\bibitem[{{Sch{\"o}nherr} et~al.(2014){Sch{\"o}nherr}, {Schwarm}, {Falkner},
  {Becker}, {Wilms}, {Dauser}, {Wolff}, {Wolfram}, {West}, {Pottschmidt},
  {Kretschmar}, {Ferrigno}, {Klochkov}, {Nishimura}, {Kreykenbohm},
  {Caballero}, \& {Staubert}}]{hmxb1}
{Sch{\"o}nherr}~G, et~al.
\href{http://dx.doi.org/10.1051/epjconf/20136402003}{2014 in European Physical
  Journal Web of Conferences, Vol.~64, European Physical Journal Web of
  Conferences} 02003.

\bibitem[{{Shakura} et~al.(2012){Shakura}, {Postnov}, {Kochetkova}, \&
  {Hjalmarsdotter}}]{shakura2012}
{Shakura}~N, et~al.
\href{http://dx.doi.org/10.1111/j.1365-2966.2011.20026.x}{2012
  \JournalTitle{\mnras} 420 216}.

\bibitem[{{Sidoli} \& {Paizis}(2019)}]{sidoli2019}
{Sidoli}~L, \& {Paizis}~A.
2019 \JournalTitle{arXiv e-prints} arXiv:1901.01882.

\bibitem[{{Staubert} et~al.(2019){Staubert}, {Tr{\"u}mper}, {Kendziorra},
  {Klochkov}, {Postnov}, {Kretschmar}, {Pottschmidt}, {Haberl}, {Rothschild},
  {Santangelo}, {Wilms}, {Kreykenbohm}, \& {F{\"u}rst}}]{staubert}
{Staubert}~R, et~al.
\href{http://dx.doi.org/10.1051/0004-6361/201834479}{2019 \JournalTitle{\aap}
  622 A61}.

\bibitem[{{Torrej{\'o}n} et~al.(2010{\natexlab{a}}){Torrej{\'o}n},
  {Negueruela}, {Smith}, \& {Harrison}}]{torrejon2010_dist}
{Torrej{\'o}n}~JM, et~al.
\href{http://dx.doi.org/10.1051/0004-6361/200912619}{2010{\natexlab{a}}
  \JournalTitle{\aap} 510 A61}.

\bibitem[{{Torrej{\'o}n} et~al.(2010{\natexlab{b}}){Torrej{\'o}n}, {Schulz},
  {Nowak}, \& {Kallman}}]{torrejon}
{Torrej{\'o}n}~JM, et~al.
\href{http://dx.doi.org/10.1088/0004-637X/715/2/947}{2010{\natexlab{b}}
  \JournalTitle{\apj} 715 947}.

\bibitem[{{Verner} et~al.(1996){Verner}, {Ferland}, {Korista}, \&
  {Yakovlev}}]{verner1996}
{Verner}~DA, et~al.
\href{http://dx.doi.org/10.1086/177435}{1996 \JournalTitle{\apj} 465 487}.

\bibitem[{{Walter} et~al.(2015){Walter}, {Lutovinov}, {Bozzo}, \&
  {Tsygankov}}]{walter2015}
{Walter}~R, et~al.
\href{http://dx.doi.org/10.1007/s00159-015-0082-6}{2015 \JournalTitle{\aapr} 23
  2}.

\bibitem[{{Walter} \& {Zurita Heras}(2007)}]{walter2007}
{Walter}~R, \& {Zurita Heras}~J.
\href{http://dx.doi.org/10.1051/0004-6361:20078353}{2007 \JournalTitle{\aap}
  476 335}.

\bibitem[{{Walter} et~al.(2006){Walter}, {Zurita Heras}, {Bassani}, {Bazzano},
  {Bodaghee}, {Dean}, {Dubath}, {Parmar}, {Renaud}, \& {Ubertini}}]{walter2006}
{Walter}~R, et~al.
\href{http://dx.doi.org/10.1051/0004-6361:20053719}{2006 \JournalTitle{\aap}
  453 133}.

\bibitem[{{Wilms} et~al.(2000){Wilms}, {Allen}, \& {McCray}}]{wilms00}
{Wilms}~J, et~al.
\href{http://dx.doi.org/10.1086/317016}{2000 \JournalTitle{\apj} 542 914}.

\end{thebibliography}
\end{document}